\documentstyle[11pt,epsf,emulateapj]{article}
\lefthead{F. Frontera et~al.}
\righthead{Emission properties of Gamma-Ray Bursts}
\def\S{{\em BeppoSAX\/}}

\makeatletter
\def\@cite#1#2{(#1\if@tempswa , #2\fi)}
\def\preprint{preprint}   \newif\ifPreprintMode
\ifx\preprint\revtex@genre\PreprintModetrue\else\PreprintModefalse\fi
\makeatother

\begin{document}

\title{Prompt and delayed emission properties of Gamma-Ray Bursts observed
 with BeppoSAX}

\author{F.~Frontera\altaffilmark{1,2},
L.~Amati\altaffilmark{1},
E.~Costa\altaffilmark{3},
J.M.~Muller\altaffilmark{4,5},
E.~Pian\altaffilmark{1},
L.~Piro\altaffilmark{3},
P. Soffitta\altaffilmark{3},
M.~Tavani\altaffilmark{6,7},
A.~Castro-Tirado\altaffilmark{8},
D.~Dal Fiume\altaffilmark{1},
M.~Feroci\altaffilmark{3},
J. Heise\altaffilmark{4},
N. Masetti\altaffilmark{1},
L.~Nicastro\altaffilmark{9},
M.~Orlandini\altaffilmark{1},
E.~Palazzi\altaffilmark{1},
R.~Sari\altaffilmark{10}
}

\altaffiltext{1}{Istituto Tecnologie e Studio Radiazioni Extraterrestri, 
CNR, Via Gobetti 101, 40129 Bologna, Italy}

\altaffiltext{2}{Dipartimento di Fisica, Universit\`a di Ferrara, Via Paradiso
 12, 44100 Ferrara, Italy}

\altaffiltext{3}{Istituto Astrofisica Spaziale, C.N.R., Via E. Fermi 21,
  00044 Frascati, Italy}

\altaffiltext{4}{Space Research Organization in the Netherlands,
 Sorbonnelaan 2, 3584 CA Utrecht, The Netherlands}

\altaffiltext{5}{\S\ Scientific Data Center, Via Corcolle 19, 00131 Roma,
 Italy} 

\altaffiltext{6}{Istituto Fisica Cosmica e Tecnologie Relative, C.N.R.,
  Via Bassini 15, 20133 Milano, Italy} 

\altaffiltext{7}{Columbia Astrophysics Laboratory, Columbia University, New
 York, NY 10027}

\altaffiltext{8}{Laboratorio de Astrof\`{\i}sica Espacial y F\`{\i}sica
 Fundamental, P.O. Box 50727, 28080 Madrid, Spain, and Instituto de 
Astrof\'{\i}sica de Andaluc\'{\i}a (IAA-CSIC), P.O. Box 03004, E-18080 
Granada, Spain}

\altaffiltext{9}{Istituto Fisica Cosmica e Applicazioni all'Informatica, 
C.N.R., Via U. La Malfa 153, 90146 Palermo, Italy}

\altaffiltext{10}{Theoretical Astrophysics, 130-33, California Institute
of Technology, Pasadena, CA 91125, USA}

\begin{abstract}
We investigated the spectral evolution in the 2--700 keV energy band
of Gamma-Ray Bursts (GRBs) detected by the Gamma-Ray Burst 
Monitor (GRBM) and localized with the Wide Field Cameras (WFCs) aboard the 
BeppoSAX satellite before May 1998. Most of them have been followed-up
with the Narrow Field Instruments aboard the same satellite.
In the light of these results we discuss open issues on the GRB phenomenon.
We find that the optically thin synchrotron shock model (SSM) provides an 
acceptable
representation of most of the time-resolved GRB spectra extending down 
to 2 keV,
except in the initial phases of several bursts and during the whole duration
of the quite strong GRB970111, where a low-energy photon depletion 
with respect to the thin SSM spectrum is observed. A strong and time variable
low energy cut-off, consistent with absorption effect, is observed during
the prompt emission of GRB980329.
We find that the X-ray afterglow starts at about 50\% of the GRB duration,
and that its fluence, as computed from the WFC light curve, is consistent
with the decay law found from the afterglow NFI observations.
We also investigate the hydrodynamical evolution of the GRB in our sample
and their associated afterglow, when it was detected. We find that the photon
index of the latest spectrum of the GRB prompt emission is correlated with
the index of the afterglow fading law, when available, as expected on the 
basis of an external shock of a relativistic fireball. We also find that for 
most of the GRBs in our sample the late emission
is consistent with a slow cooling of the  shock. Adiabatic shocks appear
more likely than radiative shocks. Parameters of the shocks at earliest times
have been derived.

\end{abstract}

\keywords{gamma rays: bursts --- gamma rays: observations --- X--rays: general
--- hydrodynamics: --- shock waves}

\section{Introduction}
In the last two years a big step forward has been accomplished in the
GRB astronomy, thanks to the BeppoSAX\cite{Boella97a} 
capability of precisely positioning these events soon after their
detection \cite{Piro98a}.  The ensuing discovery of GRB afterglow emission 
in the X-ray, optical and radio  bands
has provided the distance for some GRBs and important information on the GRB 
remnants and their environments, necessary to improve our understanding of the 
radiation mechanisms and the origin of GRBs.
\\
Time averaged spectra of GRBs seem to be consistent with the 
synchrotron shock model \cite{Tavani96}, while Inverse Compton emission 
is expected to operate at very early times \cite{Waxman97}.
\\
Models of GRBs and afterglow emission are mainly based on mechanisms of
dissipation of kinetic energy of a relativistic expanding fireball
\cite{Meszaros97,Wijers97,Vietri97}. 
There are open issues concerning the production process of the 
prompt $\gamma$-ray radiation (e.g., internal or external shocks of the 
fireball material, Kobayashi et al. 1997\nocite{Kobayashi97}),  
the time when the afterglow sets in as compared to the end of the prompt
GRB emission \cite{Sari97}, the production process of the delayed radiation 
(e.g., radiative shock vs. adiabatic shock, see Sari, Piran and Narayan 1998
\nocite{Sari98}). All these issues require observations.

BeppoSAX  offers the possibility to perform in a broad energy band 
(2--700~keV) temporal and spectral studies of the primary events from
which afterglow emission has been detected.  Indeed, when an event is 
simultaneously detected by the Gamma-Ray Burst Monitor (GRBM, 40--700~keV)
\cite{Frontera97,Feroci97} and by one of the two Wide Field Cameras 
(WFCs, 2-26~keV, Jager et al. 1997 \nocite{Jager97}), it is possible to
obtain, besides a  
precise localization (few arcmin radius error boxes), also X- and $\gamma$-ray
spectra and time profiles  of the GRBs. 
Therefore it is possible to study the spectral evolution of GRBs 
and relate the 
spectral and temporal properties of the main events with those of
the associated X--ray afterglows.
\\
Results on the spectral evolution of GRBs in a broad band were
reported by Strohmayer et al.~1998 \nocite{Strohmayer98}, using the
GRB detections obtained with the Gamma-Ray Burst Detector on board
the Ginga satellite. However,
due to the lack of knowledge of the GRB direction and thus a poor knowledge
of the instrument response function, those results were unavoidably of
limited accuracy.  
\\
We have already reported on  the time averaged spectra of some 
GRBs observed with the BeppoSAX GRBM and WFCs \cite{Frontera98b}.
In this paper we will concentrate mainly on the evolution of the
spectral properties of a sample of GRBs occurred between July 20, 1996 and 
April 25, 1998.
For completeness we include in our sample also GRB960720 and 
GRB970228, the spectral evolution of which was already studied with results 
consistent with those obtained with the present analysis 
\cite{Piro98a,Frontera98a}.
Results of comparative spectral analysis of the GRBs detected after 
April 25, 1998, will be the subject of a future paper.

\section{Instrumentation and GRB sample}

The GRBM consists of the 4 anti-coincidence shields of the Phoswich Detection 
System instrument \cite{Frontera97,Feroci97,Amati97}. Each shield 
is a CsI(Na) scintillator slab
10 mm thick with a geometric area of 1136~cm$^2$. 
Each slab is open to the sky, except along some directions, where material 
from other BeppoSAX instruments heavily decreases the transparency of the
slab $\gamma$-ray entrance window. The GRBM detector  operates 
in the 40-700~keV energy band. Two of the four GRBM units are co-aligned 
with the WFCs: unit 1 with WFC No.~1 and unit 3 with WFC No.~2.
The data available from GRBM for spectral analysis  include two 1~s
ratemeters
(40--700~keV and $>$100~keV) and 240 channel spectra in the 40--700~keV band
integrated over 128~s. For the GRB spectral evolution  we use the 1~s
ratemeters, but we check their consistency with the  GRB time
averaged spectra obtained from the 240 channel data\cite{Amati99}.
The energy resolution of  GRBM units 1 and 3 with energy has been 
discussed by Amati et al. (1997) \nocite{Amati97}; e.g., at 279~keV 
($^{203}$Hg line) it is 20\%.
The  on-axis effective area of the GRBM units 1 and 3 is 420~cm$^2$ in 
the 40--700~keV band and is 500~cm$^2$ at 300~keV.
\\
The WFC instrument consists of two coded aperture cameras, each with a field
of view of 40$^\circ \times 40^\circ$ full width at zero response and
an angular resolution of 5 arcmin.
WFCs have an energy resolution $\approx$ 20\% at 6~keV, are operated in 
normal mode with 31 channels in 2--26~keV and 0.5~ms time resolution.
The on-axis effective area of WFCs No.~1 and 2 averaged in the 2--26~keV
energy band is 118~cm$^2$ \cite{Jager97}.
\\
The follow-up observations of the GRB error boxes provided by the WFCs are 
performed with the Narrow Field Instruments (NFIs) aboard BeppoSAX, that
are orthogonal to both the WFC and GRBM axes.
They include two imaging instruments (LECS, 0.1-10~keV, Parmar et al. 1997;
\nocite{Parmar97} MECS, 2-10~keV, Boella et al. 1997b \nocite{Boella97b})
 and two direct-viewing detectors
(HPGSPC, 3-60~keV, Manzo et al. 1997 \nocite{Manzo97}; PDS, 15-300~keV, 
Frontera et al. 1997 \nocite{Frontera97}).
\\

The list of the BeppoSAX GRB events included in our sample is given in 
Table~1 along with some
basic information: position error radius, offset angle of the GRBs 
with respect to the instruments (GRBM and WFC) axis, peak fluxes and
time durations T$_X$ and T$_\gamma$ in the
X--ray (2-26~keV) and $\gamma$--ray (40-700~keV) band, respectively,
time delay
of the first NFI observation from the primary event. Positive or uncertain
(indicated by a question mark) X--ray, optical or radio afterglow detection
is reported in the last column. The GRB time durations 
are estimated from the 1~s ratemeters and give the time interval during 
which the GRB count rate is higher than the estimated background level by at 
least 2$\sigma$. As can be seen, T$_X$ is generally greater or at most equal 
to T$_\gamma$.  

\section{Spectral analysis}

The light curves from the WFCs and GRBM instruments  are shown in fig.~1. 
Each light curve  was divided into a given number of temporal
sections (see fig.~1), and a spectral analysis in the 2--700~keV energy
band was performed
for the average spectrum of each section. The duration of the sections was
chosen to be shorter during the rise of the burst in order to study the
evolution of the primary event at the earliest times.  
The spectral analysis was performed following the same procedure used
 for GRB970228 \cite{Frontera98a}. The background level
was subtracted from the GRB count rates (see fig. 1) as follows. 
For the GRBM spectra, this level was estimated using the count
rates immediately before and after the GRBs. If the background is
variable during the GRB, it is  estimated by interpolation, using a
quadratic function that fits  150~s count rate data  before the burst
and 150~s data after its end. 
For the WFC spectra and light curves, the background level is estimated 
using an equivalent section of the detector area not illuminated by the
burst or other known X-ray sources. We also checked the consistency
of this background level with that obtained by using the data before and
after the burst.
The response function of GRBM units 1 and 3 \cite{Amati97} is derived
from on-ground calibrations and checked with the Crab Nebula that was clearly 
detected
using the Earth occultation technique \cite{Guidorzi98}. The response function
is currently known with an uncertainty of about 10\%
for incident photons with low offset angles with respect to the instrument
axes, as it is the case of the GRBs in our sample (see Table~1).
We added in quadrature this uncertainty to the Poissonian
variance.
\\
We  used XSPEC software package, issue 10 \cite{Arnaud96} to deconvolve
count rate spectra, assuming a given theoretical model as input function.
In the following, all quoted errors correspond to 90\% confidence level for
each interesting parameter.
\\
A simultaneous fit to the time averaged WFC and GRBM spectra was
performed by using as input model a photo-electrically
absorbed \cite{Morrison83} smoothly broken power law \cite{Band93} given
by: \\
{\small
$$N(E) =  
       A exp (-\sigma N_H)\left(\frac{E}{100keV}\right)^{\Gamma_X} 
        \exp{\left({-E/E_{0}}\right)}$$ \\  
if $$({\Gamma_X} - {\Gamma_\gamma})\cdot E_{0} \hbox{  }\ge\hbox{  } E$$ \\ 
and $$N(E) = A \left[\frac{({\Gamma_X} - {\Gamma_\gamma}) E_{0}}{100keV}
\right]^{\Gamma_X - \Gamma_\gamma} \exp{(\Gamma_\gamma - \Gamma_X)}
\cdot \left(\frac{E}{100keV}\right)^{\Gamma_\gamma}$$ \\ 
if $$({\Gamma_X} - {\Gamma_\gamma}) \cdot E_{0}\hbox{  }\le \hbox{  } E$$
}

where $\sigma\,=\,\sigma (E)$ is the photo-electric cross-section of a gas
with cosmic abundance \cite{Morrison83},  N$_H$ is the equivalent hydrogen 
column density to the GRB,  $\Gamma_X$ and $\Gamma_\gamma$ are the power law
low energy (below E$_0$) and high energy (above E$_0$) photon indices, 
respectively, and A is the normalization parameter.
\\
The derived best fit parameters, N$_H$, $\Gamma_X$, $\Gamma_\gamma$, peak 
energy E$_p$ of the logarithmic power  ($\nu$F($\nu$)) per photon energy 
decade, are
shown  in Table 2 for each of the temporal sections. The value of
E$_p$ is given by E$_p \,=\, E_0 (2+\Gamma_X)$, under the condition that
$\Gamma_\gamma < -2$. The reduced $\chi^2$ values obtained from the best 
fits were always acceptable (less than 1.1).
We also derived lower limits on E$_p$ when  
the $\nu$F($\nu$) spectrum showed a bending at the $\gamma$-ray energies, 
but with $\Gamma_\gamma >-2$.
In this case, E$_p$ was derived under the condition that the fit to the data 
was still acceptable, assuming a $\Gamma_\gamma$ slightly lower than $-2$ 
(=~$-2.1)$.
The lower limits on E$_p$ reported in  Table~2 were thus obtained from the 
lower limits on E$_0$ corresponding to a 90\% confidence level.  
\\
When the  N$_H$ values were not constrained from the data, they were fixed 
at the Galactic values along the GRB directions, except for GRB980329, for
which we adopted the value derived from the late afterglow observation
 \cite{Zand98}.
\\
Also reported in Table 2 are the spectral and temporal properties of the
associated X--ray afterglows in the 0.1--10~keV energy band as
derived from the TOO observations with the NFIs. The two spectral parameters
(photon index $\Gamma_{a}$ and column density N$_H$) were derived assuming
a photo-electrically absorbed\cite{Morrison83} power law model with photon
index $\Gamma_a$,
while the parameter $\delta$ is the index of the fading law
I(t)$\propto t^{-\delta}$, which  best fits the 2--10~keV light 
curve.

\section{Results}

For three GRBs (GRB960720, GRB970508, GRB980425) the $\gamma$-ray emission 
starts earlier than the X-ray emission, while in the other GRBs 
$\gamma$-rays and X-rays rise simultaneously (fig.~1). Duration and  
shape of the GRB time profiles change from one GRB 
to the other, even if some similarity is observed between GRB980329 and
GRB980425, while GRB960720 and GRB970508 exhibit strikingly similar light 
curves. No X-ray precursor activity is 
detected, similar to that seen in other GRBs by the Gamma Burst Detector 
(GBD) on-board Ginga \cite{Murakami91} or WATCH/Granat \cite{Castro-Tirado94}. 
\\ 
Time duration and  shape of the GRB time profiles change from one GRB 
to the other, even if some similarity is observed between GRB980329 and
GRB980425.
\\
From the low values ($<$1.1) of the reduced $\chi^2$, we can infer that 
the GRB spectra  of our sample can be described by a 
Band law (see eq.~1), even if, in some temporal sections 
(in the case of GRB970402 in both sections), they are single power 
laws.
GRB spectra evolve with energy and with time (see Table~2).
The spectra are generally harder at low energies or they exhibit the same 
slope at  high and low energies. We 
do not find bursts  with low energy excesses with respect to the Band law 
(see eq.~1) as claimed by other authors \cite{Preece96,Strohmayer98}. 
\\
In the first few seconds, during the GRB rise (fig.~1) we
find, for 4 GRBs, that the low-energy photon index  $\Gamma_X$ is 
significantly larger than -2/3, limit slope of the low energy tail of the
instantaneous energy spectrum of an optically thin synchrotron emission 
(see Section \ref{syncr}). This feature was already noticed for GRB960720
\cite{Piro98a}. In one case (GRB970111) it holds for the entire GRB duration.
\\
The spectra are generally softer during the tail of the GRBs. A remarkable 
exception is observed in the spectrum of GRB970228, that shows a softening 
during the first pulse, but it becomes harder during the subsequent minor 
pulses of the event tail (see fig.~1). This feature was already discussed
by Frontera et al. (1998a) \nocite{Frontera98a}.
\\
If we compare the GRB photon index $\Gamma_X$ during the GRB tail 
with the photon index $\Gamma_{a}$ of the power law spectrum of the associated
X-ray afterglow, when available, it can be seen that  the afterglow 
spectrum  is generally softer than the GRB tail spectrum.
\\
Generally the absorption column densities are
consistent with the Galactic values along the GRB directions, except for
GRB980329. In this case, the N$_H$ values are higher than the Galactic
value (0.94$\times$ 10$^{21}$\,cm$^{-2}$) in the temporal sections from
A to F, while in the last two temporal sections (G and H), we are not
capable to constrain the N$_H$ value, that was frozen  to the value 
estimated during the late afterglow observation \cite{Zand98}, but
was also consistent with the Galactic value.
A much higher hydrogen column density (about 1$\times$ 10$^{23}$\,cm$^{-2}$) 
is specially apparent during the section E corresponding to
the early decay of the primary event (see fig.~1 and Table~2). 
\\
The $\nu$F($\nu$) spectrum in each of the temporal sections in which we 
subdivided the time profiles  of the GRBs in our sample is  shown in 
fig. 2.
The evolution of the peak energy E$_p$ is apparent for most of the GRBs 
studied. As an example,
GRB970228 shows a peak energy quickly crossing the 2--700~keV passband 
during the first peak, while the peak energy of GRB970111 is still within
that energy range until the
end of the event. There is no  evidence of softening in the spectra of 
GRB970402, but from the slope of the $\nu$F($\nu$) spectrum (see fig.~2), 
it is clear that E$_p$ is above 700~keV. GRB980425, presumably associated
with the supernova SN 1998bw \cite{Galama98b}, does not show special features 
with respect to the other GRBs in our samples: its peak energy decreases with
time from the GRB onset.  
\\
The GRB970111 behavior deserves particular attention. 
The X-ray spectral shape is unchanged, while the  $\gamma$-ray spectrum 
steepens monotonically. This indicates a progressive decrease of E$_p$, 
which however almost stabilizes at 30-40~keV starting about 23~s after 
the GRB onset.
This result is better apparent in fig.~3, that shows the time behavior of 
$\Gamma_X$ as a function of $\Gamma_\gamma$  for all GRBs
considered. When the peak energy is above or below the 2--700~keV 
passband, we would expect to measure approximately 
similar photon  indices $\Gamma_X$ and $\Gamma_\gamma$ (see Section
\ref{evolution}) and the value of ($\Gamma_X$, $\Gamma_\gamma$) should  
be along the diagonal (dashed line in fig. 3). On the contrary, 
$\Gamma_X$ and $\Gamma_\gamma$ could be markedly different (with $\Gamma_X$ 
close to the limit value -2/3 for an optically thin synchrotron spectrum) 
when E$_p$ is in the 2--700~keV passband. In this case 
($\Gamma_X$, $\Gamma_\gamma$) should lie out of the diagonal. If E$_p$, during
the primary event, sweeps the above passband, the plot of 
($\Gamma_X$, $\Gamma_\gamma$) should start on the diagonal, go away from it
and go back on the diagonal.
In the case of GRB970111  E$_p$ enters 
the 2--700~keV passband soon after the onset, but does not completely cross 
it until the end of the GRB. In some cases (GRB960720, GRB970228) it 
sweeps 2--700~keV passband, while  
in other cases (GRB971214, GRB980329, GRB980425) the peak energy is still in 
the passband at the end of the primary event. In the case of GRB970402, from
the rising of its $\nu$F{$\nu$ spectrum (fig.~2),
the peak energy is clearly above 700~keV,  while for GRB970508 there 
is a hint that E$_p$ crosses the 2--700~keV passband.
\\
In only one case (GRB971214) we observe a hardening in the afterglow spectrum 
with respect to that observed during the prompt emission as expected in the 
model discussed by Waxman (1997) \nocite{Waxman97}.

\section{Discussion}
\label{disc}
We discuss our results in the light of current theories on GRBs in order to
contribute to solve many open issues concerning the GRB phenomenon.

\subsection{Test of the Synchrotron shock model}
\label{syncr}
As discussed by several authors (see, e.g., Katz 1994, Tavani 1996, Sari \&
Piran 1997b)\nocite{Katz94,Tavani96,Sari97b}, the electromagnetic 
radiation from GRBs and their afterglows is likely due to synchrotron  
radiation from accelerated 
particles in internal or external shocks of a relativistically expanding 
fireball (Synchrotron Shock Model, SSM). The observed hard-to-soft change of 
the spectrum with energy 
and with time from the GRB onset is in agreement with this picture
\cite{Dermer99}. 
\\
Assuming that the emission region is optically thin, the expected power law 
index of the  $\nu$F($\nu$) spectrum below the peak energy E$_p$ 
is expected to be in the range from 1/2 to 4/3 (corresponding to -3/2 to -2/3, 
respectively, in the photon spectrum), where the first index is obtained when
cooling of the particle distribution during the GRB is also taken into 
account, while the latter index is reached in the case of an instantaneous
spectrum or for monochromatic particles \cite{Cohen97}.  Notice that the limit
of -2/3 is independent of the electron distribution or uniformity of the 
associated magnetic field.
\\
As discussed by Tavani (1996) \nocite{Tavani96}, the Band law is a 
good approximation of the spectrum expected by the thin SSM and our results 
confirm this fact.
However, using this simple model, we find that, for 50\% of the GRBs 
(see fig.~4), $\Gamma_X$ is above the expected limit photon index (-2/3) in 
the earliest time intervals of the light curve. In the case of the intense 
GRB970111, this holds almost for the entire GRB duration.
Crider et al. (1997) \nocite{Crider97} found a similar violation in 
the 30-1800~keV spectra of GRBs observed  by BATSE, while Strohmayer 
et al. (1998) and Preece et al. (1998) \nocite{Strohmayer98,Preece98} find 
this violation in time averaged spectra of  GRBs detected with the GBD/Ginga 
and BATSE/CGRO experiments, respectively. 
\\
Our data confirm that some 
additional mechanism is certainly active for some GRBs, that modifies 
the optically thin synchrotron spectrum at very early times.
This could be synchrotron self-absorption\cite{Hara99}, Compton up-scattering 
of low-energy photons \cite{Liang97} or plasma physics effects \cite{Tavani99}.
Assuming single Compton up-scattering by highly relativistic electrons 
with random  direction, the 
limit photon index of the low-energy spectrum is zero \cite{Rybicki79}. 
We find that the low-energy photon index of all GRBs in our sample,  except 
that of GRB970111, is consistent with this  limit value.
\\
The approximate stabilization  of E$_p$ at $\sim$35~keV for GRB970111,  
corresponding to a decrease of the power law photon index $\Gamma_\gamma$ 
will be discussed in Section \ref{970111}. 
For the other GRBs, the change of the photon index  
observed starting from the GRB tail up to the late X-ray afterglow 
observations not as much relevant.
\\
The high N$_H$ observed during the prompt emission of GRB980329 and its
decrease during the tail of the event can be interpreted in two ways.
It might be  a consequence of variable absorption of external
intervening material due to its progressive  photo-ioniziation as the GRB 
evolves \cite{Bottcher98}. In this case B\H{o}ttcher et~al.\ (1998) 
\nocite{Bottcher98} also expect fluorescence features and/or K edges 
that are not apparent in the prompt and delayed
emission \cite{Zand98}. However 
these features could be too weak to be detectable with our instruments and/or
could be not within the WFC or NFI passband, due to the redshift.
The redshift of GRB980329 is still unknown (see a possible estimate in
Fruchter 1999) \nocite{Fruchter99}. The other possibility is
that intrinsic absorption takes place at the GRB site. If this is the case, 
in the context of the fireball shock model \cite{Sari98},  N$_H$ 
is given by E/($\gamma_0$m$_p$c$^2$) divided by 4$\pi$$r^2$,
where E is the energy of the shock, $\gamma_0$ is initial
Lorentz factor of the shocked material, m$_p$ is the proton mass, c is the
light velocity ad r is the distance from the explosion center to the shock 
dissipation radius. With typical values for the energy
E ($\sim 10^{52}$~erg) and $\gamma_0$ ($\sim 100$), the measured N$_H$ gives
a value of r$\sim 10^{14}$~cm, which is inconsistent with external shocks 
that are expected only
at values of R$\sim 10^{16}--10^{17}$~cm, when the N$_H$ would be lower by a 
factor of $10^4-10^6$.
Thus our results points rather to internal shocks accompanied by rapid 
expansion of a fireball.

\subsection{Earliest X-ray afterglow emission}
\label{afterglow}
Table 3 shows the X-ray (2-10~keV) and $\gamma$-ray (40-700~keV) fluences,
S$_X$ and S$_\gamma$, of the GRBs in our sample as
derived from our spectral analysis. For comparison, when available, also
the 2-10 keV fluence  S$_{a}$ of the afterglow emission is given.
The latter was estimated from the fading laws (see Table~2) integrated
over the time interval from the GRB end up to 10$^6$~s, when the X-ray 
afterglow emission level is generally already negligible for a power 
slope $\delta$ of the fading law greater than 1, as in our case.
The ratio S$_X/S_\gamma$ changes from event to event
and ranges from 1\% to about 40\%. It does not seem that detection of X-ray 
afterglow emission is related to the amplitude of the above ratio.
Also the ratio S$_a/S_\gamma$ changes from a GRB to another, ranges
from 0.4\% to about 20\% and does not appear to be strictly related with the 
S$_X/S_\gamma$ 
value, as also evident from the S$_a/S_X$ ratio. This ratio also
shows that the fluence of the X-ray afterglow emission is, within
a factor 2, of the same order of that measured, in the same energy band,
during the primary event.
\\
In the scenario of the fireball model,
in order to explain the complex time profiles of GRBs, it was shown
\cite{Kobayashi97,Sari97b} that the GRB main event could be due to 
internal shocks, although for simple fast-rise/exponential-decay (FRED) events
also external shocks can be accepted \cite{Dermer99}.
Instead, there is general consensus that the GRB late afterglow emission
is likely due to the external shocks propagating in the interstellar
medium \cite{Meszaros97,Sari98}. In this 
scenario, the two phenomena can evolve in different ways and can involve
energetics not necessarily correlated. However it has been suggested
\cite{Sari97,Sari99b} that the early afterglow could start few dozens of
seconds after the burst. We tested this prediction  for the GRBs 
in our sample.
We have already shown (see, e.g., Costa et al. 1998 and 
Piro et al. 1998 \nocite{Costa97,Piro98b}) that the extrapolation of 
the 2-10~keV afterglow fading law back to the time of the GRB is in agreement 
with the 2-10~keV flux measured during the last portion
of the event. This should suggest that at least the tail of
the GRBs could be due to afterglow emission, while at the beginning
of the primary event, the contribution to the GRB emission from the 
afterglow is negligible.
We adopt the following procedure to perform the test.
It is reasonable to assume that the intensity of the X-ray afterglow depends 
on the spectral index
$\Gamma_a$ and temporal index $\delta$ according to the following law:
      \begin{equation}
                      I(E,t)\, \propto \, E^{\Gamma_a} t^{-\delta}
      \end{equation}
With this assumption, the more general expression of the ratio between 
the 2--10~keV fluence in the time intervals (t$_1$, t$_2$) and 
(t$_2$, $\infty$), with t$_2>$t$_1$, is given by
   \begin{equation}
               R(t_1,t_2)\,\equiv \, S(t_1,t_2)/S(t_2,\infty) \,= \,
               K(\bar{\Gamma}_1,\bar{\Gamma}_2)[(t_1/t_2)^{-\delta +1} - 1]
    \end{equation}
where S(t$_1$,t$_2$) is the 2-10~keV fluence in the (t$_1$,t$_2$) interval,
S(t$_2$,$\infty$) is the fluence in the interval from $t_2$ to a time 
t $\gg$ t$_2$
where the afterglow intensity has decreased to a negligible value, 
$\bar{\Gamma}_1$ and $\bar{\Gamma}_2$ are the average photon indices of the 
afterglow power law spectral
emission  in  (t$_1$, t$_2$) and  (t$_2$, $\infty$), respectively.
Notice that parameter $K(\bar{\Gamma}_1,\bar{\Gamma}_2)$ must be
greater than 1 if the spectrum becomes softer with time as observed
(see Table~2).
\\
If we express t$_1$ and t$_2$ in terms of the time duration T of each GRB,
that is, t$_1$~=~f$_1$T and t$_2$~=f$_2$T, with f$_1<$f$_2$, the previous
equation becomes
   \begin{equation}
     R(f_1,f_2) \,= \,
               K(\bar{\Gamma}_1,\bar{\Gamma}_2)[(f_1/f_2)^{-\delta +1} - 1]
               \label{formula}
   \end{equation}

The ratio R(f$_1$,f$_2$) for each GRB of our sample was evaluated 
assuming  f$_2\,=\,1$ and different values of f$_1$. The 
result obtained for f$_1\,=\,0.01$ and  f$_1\,=\,0.63$ is shown in fig.~5. 
The above law  gives
a good fit to the data ($\chi^2_{\nu}=1.04$, dof=4) in the case 
f$_1\,=\,0.63$, but an 
unacceptable fit ($\chi^2_{\nu}=2.5$, 4 dof) in the case f$_1\,=\,0.01$ when
the total GRB fluence is used. We also checked the behavior of the above
ratio  in the case f$_1\,=\,0.40$, obtaining a reduced $\chi^2_{\nu}=1.3$.  
In fig.~5 (bottom panel) we show the best fit 
curve, with K~=~1.66$\pm0.36$, in the former case. For comparison, the 
curve with K=1, under the assumption  of no spectral evolution from 
(t$_1$, t$_2$) to (t$_2$, $\infty$), is also shown. In fig.~5 (top panel), 
we report the 
expected curve with the minimum value of K allowed (K=1) in the case 
f$_1 \,=\,0.01$. There is no agreement between data and expected dependence 
of R(0.01,1) on $\delta$.
\\
This result shows that roughly the second half of the GRB X-ray light curve 
is consistent with afterglow emission.
Likely GRB time profiles are the 
superposition of two components, one due to the process that produces 
the main event and the other due to the delayed emission.
A possible scenario envisages the former component as due to internal shocks
and the latter due an external shock (see, e.g., Sari (1997)\nocite{Sari97}).
If the emission due to internal shocks has a short duration, the afterglow 
can be separated from the internal component, that gives rise to the earliest
emission of a GRB.  
This could be the case of GRB970228\cite{Costa97,Frontera98a}, in which the 
first peak 
(see fig.~1) could be due to an internal shock, while the late emission
could be due to external shocks and thus coincide with the afterglow emission.
\\
According to the internal-external shock model for the GRB,
 from the above results, it is possible to evaluate the initial Lorentz 
factor $\gamma_0$ of the shocked fluid \cite{Sari99a,Sari99b}:
   \begin{equation}
                  \gamma_0\,=\,240 E^{1/8}_{52} n_1^{1/8} (t/10~s)^{-3/8}
   \end{equation}
where t is the onset time of the afterglow emission, E$_{52}$ is the 
afterglow energy released in the shock in units of 
10$^{52}$~erg and n$_1$ is the number density of the ambient medium in units
of cm$^{-3}$. If the early afterglow starts at about 50\% of the 
GRB duration, for those GRBs for which it is possible to make an estimate
of E$_{52}$ (see Table~3), we obtain the values of
$\gamma_0$ reported in Table~4, assuming a constant ambient medium density 
n$_1 \,=\,1$. The luminosity distance was evaluated from
the redshift values reported in Table 3, assuming
a standard Friedman cosmology with H$_0\,=\,65$~km/s/Mpc and $\Omega\,=\,0.3$.
The energy released was assumed to be isotropic.
The value of $\gamma_0$ is of the order of 150 for all GRBs in
our sample with known distance, except for GRB980425. For this peculiar GRB, 
we have assumed its coincidence with the type Ic supernova SN1998bw 
\cite{Galama98b}. In this case,
Pian et al. 1999b\nocite{Pian99b} estimate an afterglow emission with fading
law index $\delta > 1.3$. This upper limit was obtained assuming that the 
afterglow starts during the GRB tail as discussed above, and considering only 
the first TOO measurement as main contribution to the afterglow radiation, 
while the radiation detected about 6 months after the main event was assumed 
to be mainly due to 
thermal emission from the supernova. The much lower value of 
$\gamma_0$ ($<50$) obtained for this GRB shows that the association of 
GRB980425 with SN1998bw implies a deviation from the extreme
relativistic conditions expected for GRBs in the context of the fireball 
scenario and thus that it could be a member of another class of GRBs. 
Assuming that this event is not
associated with SN1998bw, but with a fireball with redshift z~$\sim$~1, we 
could obtain a value of $\gamma_0$ consistent with the values found for the 
other GRBs.

\subsection{Hydrodynamical evolution of a shock}
\label{evolution}
An observational test of the hydrodynamical evolution of GRB primary events 
and afterglows is of key importance to understand the GRB 
phenomenon. The afterglows evolution resulting from forward shocks
with the interstellar medium (ISM) was theoretically investigated
by several authors\cite{Sari98,Meszaros97,Meszaros99}, while the GRB
evolution is less clear, except the case of FRED (fast rise, exponential
decay) events\cite{Dermer99}. Here we  test the expectations of the
model by Sari, Piran and Narayan (1998) \nocite{Sari98} making use of our spectral 
results. This model considers a spherical shock that propagates in a
surrounding medium (interstellar medium, ISM) of constant density. The
electrons accelerated in the shock are assumed to have a power law
distribution of Lorentz factor $\gamma_e$ (N($\gamma_e)\propto \gamma_e^{-p}$)
where $\gamma_e\, = \, E_e/m_e c^2 \ge \gamma_m$  and electron 
distribution index $p>2$. Two extreme cases of hydrodynamical evolution
of the shock are considered: a fully adiabatic and a fully radiative
evolution. In the case of an adiabatic  evolution the internal energy
of the shock is constant, while in the other case it varies with the bulk
Lorentz factor $\gamma$. The time dependence of  $\gamma$ is
different in the two regimes.
\subsubsection{Late GRB spectrum vs. afterglow decay}
\label{decay}
If the GRB tail is due to afterglow emission and thus to an external shock, 
we should also expect
a correlation between spectral properties of the GRB tail
and power-law index $\delta$ of the fading afterglow emission. 
The model by Sari, Piran and Narayan (1998) \nocite{Sari98} predicts the
occurrence of two breaks in the synchrotron spectrum, one at energy E$_m$
corresponding to the lowest energy of the injected
electrons with Lorentz factor $\gamma_m$, and the other at energy E$_c$ 
(cooling break),
corresponding to the electron energy with Lorentz factor $\gamma_c$,
above which cooling for synchrotron radiation is significant. 
Figure 6 shows the expected evolution of the $\nu$F($\nu$) spectrum for fully
adiabatic and fully radiative cooling of the shocks, for typical 
values of the model parameters (see values in the figure caption). 
Due to the different variation with time of E$_c$ and E$_m$ 
(see Sect.\ref{peak}), at early times, E$_m >$E$_c$  (fast cooling), while
at late times, E$_m < $E$_c$ (slow cooling).
The passage from fast cooling to slow cooling occurs at a time t$_0$ 
when E$_m$~=~ E$_c$~=~E$_0$.  It was shown
\cite{Sari98} (see also fig.~6) that the expected  power law photon index 
of the afterglow spectrum is given by -3/2 for E$_c <$E$<$E$_m$ and by -p/2-1 
in the high energy tail (E higher than E$_m$ and E$_c$), independently of the 
cooling mechanism of the shock (radiative or adiabatic).
From Fig.~2 and its comparison with fig.~6, we can see that, in the case of 
GRB970228 and GRB970402, the high energy photon index $\Gamma_\gamma^l$ of 
the last section of
the time profiles is consistent with -3/2, while for the other GRBs in
our sample it is likely related with the high energy tail of the associated
afterglow.
We make this assumption setting, for all GRBs in our sample except 
GRB970228 and GRB970402, $\Gamma_\gamma^l$~=~ -p/2-1.
\\
The fading law of the 2-10~keV afterglow emission  
(I(t)$\propto$t$^{-\delta}$)  provides a further relation between p and 
$\delta$.
Indeed the model by Sari, Piran and Narayan (1998) \nocite{Sari98} expects 
that the index $\delta$ is related to p through 
$\delta\,=\,3p/4-1/2$ in the case of an adiabatic cooling of the shock 
when both E$_m$ and E$_c$ are below the 2--700~keV passband at the epoch t 
of the afterglow observation. This is likely our case.
The same model expects that, in the case of a radiative cooling of the shock 
(in this case the shock becomes adiabatic at time t$_0$ when  E$_m$~=~ E$_c$),
$\delta\,=\,6p/7-2/7$.
Combining all the above relations, we found the following dependence
of $\Gamma_\gamma^l$ on $\delta$: $\Gamma_\gamma^l\,=\, -4/3-2/3 \delta$
for an adiabatic shock, and $\Gamma_\gamma^l\,=\, -7/6-7/12 \delta$, for a 
radiative shock.
\\
The measured values of $\Gamma_{\gamma}^l$  as a function of $\delta$
are shown in fig.~7a. Superposed to the data are the curves derived from the
above relations. For comparison, we show in Fig.~7b, for all GRBs in our 
sample for which the afterglow spectrum is known, the power-law photon
index $\Gamma$$_a$ of the X-ray afterglow (see Table~2) as a function of the
corresponding index $\delta$ of the fading law  in the same energy band.
Also shown in the figure are the expected
$\Gamma_a$ vs. $\delta$ relationships \cite{Sari98}:
$\Gamma_a\,=\, -7/6-7/12 \delta$ for radiative cooling (t$<$t$_0$), while
for an adiabatic cooling, $\Gamma_a \,=\, -2/3 \delta - 1$ or
$\Gamma_a \,=\, -2/3 \delta - 4/3$, depending on the observing time t after
the primary event, if t$_m<$t$<$t$_c$ or t$>$t$_c$, respectively.
The parameters t$_m$ and t$_c$ are the times at which E$_m$ and E$_c$  
sweep the energy passband of the MECS detectors, respectively (see fig.~6).
\\
From fig.~7a, it appears that the data are consistent with
the expected dependence of $\Gamma_\gamma$ on $\delta$, but their
uncertainties
do not allow us to discriminate among different cooling models.
The peculiarities of GRB970111 and GRB980425 are also apparent from this plot.
Assuming that the afterglows associated with these GRBs satisfy one of
the two plotted relationships, the X-ray afterglow of GRB970111 should 
have exhibited a fading law with power-law index lower than -2, that 
is consistent with its non detection with BeppoSAX 17hrs after the primary 
event. 
The same condition applied to the GRB980425 afterglow implies a 
much lower $\delta$ than -2. 
\\
Also the measured $\Gamma_a$ vs. $\delta$ are consistent with the expected 
relationships between these two quantities derived from the synchrotron shock 
model\cite{Sari98}, but, also in this case, due to the
uncertainties in these parameters, we cannot discriminate among 
different cooling laws.

The above discussed $\delta$ vs. p and $\Gamma$ vs. p relationships 
have also  permitted the estimate of the index of the electron 
distribution for 5 GRBs in our sample. Given the lower uncertainty 
in the value of $\delta$ (see Table~2), we report in Table~4 the values
of p derived from  $\delta$. We assume an adiabatic cooling (see sect.
\ref{peak}) in the two cases above considered (t$_m<$t$<$t$_c$, and 
t$>$t$_c$). The corresponding values of p derived from $\Gamma_a$, due to 
the larger uncertainty in this parameter (see Table~2) are consistent with 
the p values reported in Table~4. These change from one GRB to the
other and rage from 2.1 to 3.1. It is remarkable that for the two GRBs
(971214 and 980329) for which it was possible to derive the values of p 
from the spectral index $\Gamma_\gamma^l$ of the prompt 
emission, we found values of p also consistent with those reported in Table~4
(3.2$\pm$1 for GRB971214 and 2.6$\pm$0.6 for GRB980329).

\subsubsection{Temporal evolution of the peak energy E$_p$}
\label{peak}
As shown in fig.~6, the  model by Sari, Piran and Narayan (1998) 
\nocite{Sari98} 
predicts a temporal evolution of the $\nu$F($\nu$) spectrum of the GRB 
afterglow. A spectral evolution is also expected during
the primary event, even if the time  behavior can be more complex: it can 
change from one GRB to the other and can depend on the assumed model
(e.g., external shocks, Dermer, B\H{o}ttcher and Chiang 1999 \nocite{Dermer99}
vs. internal shocks, Sari 1997 \nocite{Sari97}). 
We observed both spectral evolutions (see Table~2 and fig. 2): on the basis
of the results discussed in the previous sections, the spectra obtained in 
the first half of the GRB duration are mainly due to the primary event, 
while those obtained later are likely due to afterglow emission 
(see Sect.\ref{decay}).
\\
To characterize the spectral evolution, we consider the time behavior of the 
peak energy E$_p$ of the $\nu$F($\nu$) spectrum.
The model by Sari, Piran and Narayan (1998) \nocite{Sari98} predicts the occurrence of a 
maximum in the $\nu$F($\nu$) 
spectrum in correspondence of one of the two breaks in the synchrotron 
spectrum, at energies E$_m$ and E$_c$, respectively 
(see Sect.\ref{afterglow}). 
These parameters decrease with time in a different manner and according
to the assumed cooling type of the shock (adiabatic or radiative).
At early times from the afterglow onset (t$<$t$_0$) 
the peak energy E$_p$ is coincident with $E_m$ that decreases as t$^{-3/2}$ for
adiabatic shock and as t$^{-12/7}$ for a radiative shock, while at late times 
(t$>$t$_0$) it is coincident with E$_c$ that decreases as t$^{-1/2}$ for an
adiabatic shock and as t$^{-2/7}$ for a radiative shock (see fig.~6). 
As discussed above, at the time $t_0$  E$_m$ coincides with E$_c$ 
(and with E$_p$). This time is different for an adiabatic or a radiative 
cooling (see, e.g., fig.~6). Notice that these temporal dependences of
E$_m$ and E$_c$ hold during the decay phase of the afterglow,
while at the beginning of the fireball interaction with the ISM one expects a 
shallower decay of the break energies.
\\
Thus the observed time behavior of E$_p$  can 
constrain the cooling mechanism of the shock and  provide the
value of important physical parameters, like the fraction $\epsilon_e$ of
shock energy that goes into the electrons and the fraction $\epsilon_B$ 
that goes in magnetic energy density \cite{Sari98}.   
\\
Figure~8 shows the behaviour of E$_p$ with respect to time for  all GRBs in 
our sample. The values are not corrected for GRB redshift, even when this is 
known. We have verified that this correction is not critical for our 
conclusions. 
For the GRBs with detected late afterglow, we also show in fig.~8, when 
available, the values of E$_p$  obtained from the literature
(GRB970228, Masetti et al. 1999 \nocite{Masetti99}; GRB970508, Galama et
al. 1998a \nocite{Galama98a}; GRB980329, Palazzi et al. 1998 
\nocite{Palazzi98}; GRB971214, Dal Fiume et al. 1999 \nocite{Dalfiume99},
Ramaprakash et al. 1998 \nocite{Ramaprakash98}).
Superposed to the data is also  the expected time behavior of E$_m$
for short times from GRB onset in both the radiative and adiabatic conditions 
of the shock. In addition we show the behavior of E$_c$ when afterglow emission
is detected and the spectrum is known. The E$_m$ curves
are normalized to the earliest measured values of E$_p$,
while the E$_c$ curve is normalized at the value of E$_p$ obtained from 
the late afterglow spectrum.
In fig.~8  the time at which the afterglow is expected to have already started 
(see Sect.\ref{afterglow}) is also shown (vertical dashed line). 

For most of the GRBs in our sample, only lower limits
of E$_p$ are available in the earliest part of the event. In spite of that,
it is apparent that $E_p$ decreases with time more quickly than
expected in the case of a fast cooling (adiabatic or radiative) of an
external forward shock. This is another evidence for internal shocks during
the first part of GRBs.
\\
Another important feature of the data is that, for GRB970228 and GRB970508, 
the extrapolation of the fast cooling curves at late times
is clearly inconsistent with the value of E$_p$ measured during the afterglow
 about 10 days after the primary event.
This means that a transition from a fast cooling to a slow cooling phase
of the shock has occurred in the meantime. For GRB970508 we report in fig.~8, 
in addition to the late value of E$_p \equiv$E$_c$, the estimated
value of E$_m$ \cite{Galama98a}. This  break fits 
nicely with the adiabatic evolution curve, confirming the passage of the
shock
from fast to slow cooling. The backward extrapolation of the 
expected behavior of E$_p$ normalized to the last data point allows us 
to derive the transition time t$_0$ from fast cooling to slow cooling. 
The t$_0$ values obtained are reported in Table 4, along with the rough
estimate of the same parameter for other two GRBs (GRB971214 and GRB980309), 
for which the transition time t$_0$ is more uncertain. 
\\
The E$_p$ behavior alone does not permit us to clearly establish whether
at early times the shock cooling is adiabatic or radiative. We obtained this
information by comparing the maximum of the energy spectra F$_{max}$
at early time (during the latest section of GRB time profiles) with that at 
late time (during late afterglow observations). The results are shown in 
Table~4. 
In the case of GRB970508 and GRB980329, F$_{max}$(early) and F$_{max}$(late) 
are comparable, while in the case of GRB971214, F$_{max}$(early) is a 
factor from about 4 to 18 higher than F$_{max}$(late). This range is due
to the uncertainty in the true value of F$_{max}$(late).
The two values reported in Table~4  are those given by  
Ramaprakash et al. (1998) \nocite{Ramaprakash98} and
Dal Fiume et al. (1999) \nocite{Dalfiume99}, respectively.
The difference is due to the different absorption by dust assumed by these
authors to correct the measured magnitudes.  
Dal Fiume et al. (1999) \nocite{Dalfiume99}
assumes the extinction typical of starburst galaxies 
and take into account the redshift of the measured radiation, while 
Ramaprakash et al. (1998) assume an exponential absorption with 
optical depth from extinction $\propto1/\lambda$, where $\lambda$ is the 
wavelength of the observed radiation. 
\\
As a result, at early times, for  GRB970508  and GRB980329, we see  an 
adiabatic evolution of the shock, while for  
GRB971214, the shock cooling is close to be adiabatic if we assume the highest
value of F$_{max}$(late). As discussed by  
Dal Fiume et al. (1999) \nocite{Dalfiume99}, there are reasons to prefer 
their correction, which implies that the shock is almost adiabatic.
\\
Also the  values of the shock parameters $\epsilon_e$ and $\epsilon_B$ are 
reported in Table~4. 
In the case of GRB970228, we are not able to estimate F$_{max}$(late) 
from the available data \cite{Masetti99}. For this GRB we report 
the values of these parameters for both a radiative and an adiabatic fast 
cooling.
It is noterworthy that the derived $\epsilon_B$(late) and 
$\epsilon_B$(early) are roughly in agreement and that they are  lower than  
$\epsilon_e$. This is expected when the contribution from Compton 
up-scattering cannot be neglected.  
The value of t$_0$  derived by us for GRB970508 is consistent with the 
value (about 700~s) derived by Galama 
et al.(1998a)\nocite{Galama98a}, who also noticed that an adiabatic model 
fits better the late afterglow data. 

\subsubsection{The peculiar behavior of GRB970111}
\label{970111}

We already have discussed in Sections 4 and 5.1 some peculiarities 
of GRB970111.
This GRB stands out for its low energy spectral slope, that is not consistent
with a thin synchrotron model and with Inverse Compton, making self-absorption
a possible explanation.
GRB970111 is also peculiar for its E$_p$ time behavior: a
transition in the rate of E$_p$ decrease is observed in correspondence of the 
expected starting time of the afterglow emission (see fig.~8). However this 
break cannot be interpreted as passage of the cooling break E$_c$ through
the 2-700~keV passband, since the
change of the spectral slope above and below E$_c$ is predicted
to be 1/2 and we see a much greater change. It therefore must be E$_m$ which
should however decrease as t$^{-3/2}$ in the adiabatic case or faster
(as t$^{-12/7}$) in the radiative case. Instead the slope of E$_p$ is 
consistent with t$^{-1/2}$ that is not in agreement with the decay phase
of the afterglow.
\\
The evolution of the shock does not appear to be adiabatic.
The non adiabatic character is apparent from
the time behavior of F$_{max}$. In the second half of the GRB time profile,
the energy spectrum shows an intensity maximum that decreases with time 
according to a power law with index 2.6, while a constant intensity is expected
for an adiabatic cooling. The measured slope is also inconsistent with
a radiative cooling, in which case a much lower slope (-3/7) is 
expected for  the power law time decay of F$_{max}$. The extrapolation of 
the time behavior of 
E$_p$ and F$_{max}$ at the time of the first BeppoSAX observation yields
X-ray fluxes that are well below the upper limit 
given
by Feroci et al. (1998) \nocite{Feroci98}. Also the upper limits of
the afterglow given in the optical band \cite{Gorosabel98} are well above
this extrapolation.
\\
A possible intepretation of this strange E$_p$ behavior is that we are 
observing
two successive electron acceleration episodes, giving rise to the 
first and the second pulses (see $\gamma$-ray time profile in fig.~1).
During the first pulse, E$_p$ is supposedly decaying but never enters the
2--700~keV passband (as soon as injection stops the peak can sweep this 
energy band faster than allowed to observe with our time integration of 
the spectra), while during the second pulse 
the electron distribution has a long cooling timescale that makes the peak 
almost stable at around 35 keV. This second episode (that could also be
originated by an external shock) should show a roughly constant bulk 
Lorenz factor 
that later decays and gives all the power dependences on the lightcurve.
Following this interpretation, the electron power law dependence 
on $\Gamma_\gamma$ is p= 1 -2$\Gamma_\gamma$ for the early portions of the 
second pulse and p= -2($\Gamma_\gamma$+1) for the later ones \cite{Sari98}.
From the measured values of $\Gamma_\gamma$ we find, for the second pulse
of the GRB, p$\sim$4. 
This electron index can account for the non detection  of X-ray/optical
 afterglow emission about 13~hrs after the primary event.

\section{Conclusions}
Several conclusions can be drawn from the above results.
The optically thin synchrotron model (e.g. Tavani 1996 \nocite{Tavani96}) 
appears to be consistent with the GRB spectra reported
in this paper. It can describe most of them and their evolution 
with time and energy, but not the low energy spectra at earliest times after
the GRB onset, where there is a deviation from its expectations. 
Some additional ingredient, probably self-absorption or Compton up-scattering,
is required to explain the low energy spectra at early times. 
In the case of GRB970111, the deviation is the largest and lasts for
the entire GRB duration. 
\\
In one case, GRB980329, a relevant and variable absorption has been detected
during the prompt emission. This feature appears to be
consistent with variable self-absorption of the X-ray emission produced in 
internal shocks of a relativistic expanding fireball, but absorption from
external material that progressively becomes photo-ionized by the
GRB radiation cannot be excluded.
\\
The fluence of the second half of GRB time profiles is consistent with that
expected from  afterglow emission. With this assumption, in the context of
the fireball shock model (e.g. Sari, Piran and Narayan 1998 \nocite{Sari98}), 
we have derived the initial Lorentz factor $\gamma_0$ of the shocked 
material for 5 GRBs in our sample, for which the redshift has been
determined. The value of $\gamma_0$ is about 150 for all these events, 
except for GRB980425 if this GRB is associated with the supernova SN1998bw. 
In this case we find an upper limit  $\gamma_0<50$, that is lower than the
typical values assumed in relativistic expanding fireballs. This GRB 
could be a member of another class of events.
\\
We have tested the presence of a correlation between photon index of
the late energy spectra of the GRB prompt emission and index of the fading
law of the associated afterglow. We find that this correlation exists and
it is consistent with the hydrodynamical evolution of an external shock as  
discussed in the model by Sari, Piran and Narayan (1998). \nocite{Sari98}
However the uncertainty in the data does not allow the discrimination
between different types of evolution (adiabatic vs. radiative, fast cooling vs.
slow cooling).
\\
The shock cooling properties have been investigated from
the temporal behavior of the peak energy of the $\nu$F($\nu$) spectrum. 
We have still assumed the external shock model by Sari,
Piran and Narayan (1998) \nocite{Sari98}. We find that the values of 
E$_p$ measured in the first half
of the GRB time profile have a time behavior which is not consistent
with an external shock and confirm that internal shocks have a key role in
 determining the GRB spectral evolution at earliest times.
\\ 
Comparing the values of E$_p$
in the second half of the event with those extimated
at late times ($\sim 10^6$~s from the GRB onset) when available, we find that 
for most of these events, a transition from a fast cooling to a slow cooling
has taken place. Adiabatic evolution seems to be preferred.
In the case of GRB971214, the uncertainty in the peak flux of the afterglow
spectrum does not allow the evolution to be determined, although an almost
adiabatic evolution is  preferred. 
\\
We also have determined the fractions
of the shock energy that go in electrons and magnetic energy, and the 
index p of the electron distribution accelerated in the shock. We find that p
changes from one GRB to the other and ranges from about 2.1 to 3.1, for
an adiabatic shock cooling.
\\
The case of GRB970111 merits particular consideration. From this GRB 
no afterglow emission was detected about 17~hrs after the primary event.
The GRB time profile in the 40-700~keV energy band  shows two main pulses,
with the second one that occurs in the second half of the GRB duration.
We find that the time behavior of the peak energy of the $\nu$F($\nu$) 
spectrum  
shows a break in correspondence of the rise of this pulse. If we assume
for this GRB what we found for other GRBs with known 
afterglow emission, this second pulse could be due 
to an external shock. However the time behavior of E$_p$ is not consistent
with that  expected during the decay phase of the fast cooling of a 
spherical shock. A possible
interpretation, in the context of an external shock of a fireball model, is 
that during the second pulse we are  observing the phase in which
the Lorentz factor of the shocked material is still constant or is
decreasing very slowly. The value of p found for this GRB is about 4, that
justifies its non detection  17~hrs after the primary event.

\begin{acknowledgments}
We are very grateful to Hara Papathanassiou for very useful comments
and suggestions on the manuscript. This research is supported by the 
Italian Space Agency ASI

\end{acknowledgments}
}

\clearpage

\begin{deluxetable}{lccccccccccl}
\tablewidth{0pt}
\tablenum{1}
\tablecaption{GRBs included in our sample}
\label{table}
\tablehead{
\colhead{GRB} &
\colhead{Position} &
\colhead{Offset wrt} &
\colhead{X--ray \tablenotemark{(a)}} &
\colhead{T$_{x}$} &
\colhead{$\gamma$--ray \tablenotemark{(a)}} &
\colhead{T$_{\gamma}$} &
\colhead{1$^{st}$ TOO} &
\colhead{Counterparts} \nl
   & error radius& inst. axis & peak flux & (s) & peak flux & (s) & delay &} 
\startdata
GRB960720 & 3' & 11.1$^{\circ}$ & 0.25 & 17 & 10 & 8 & 45$^{d}$ &   \nl 
GRB970111 & 3' & 14.6$^{\circ}$ & 1.4 & 60 & 56 & 43 & 16$^{h}$ & X ?  \nl 
GRB970228 & 3' & 13.1$^{\circ}$ & 1.4 & 55 & 37 & 80 & 8$^{h}$ & X, opt. \nl
GRB970402 & 3' & 8.5$^{\circ}$ & 0.16 & 150 & 3.2 & 150 & 8$^{h}$ & X  \nl 
GRB970508 & 1.9' & 10.3$^{\circ}$ & 0.35 & 29 & 5.6 & 15 & 5.7$^{h}$ & X, opt.,radio  \nl
GRB971214 & 3.3' & 16.5$^{\circ}$ & 0.2 & 35 & 6.8 & 35 & 6.7$^{h}$ & X, opt.  \nl
GRB980329 & 3' & 19.2$^{\circ}$ & 1.3 & 68 & 51 & 58 & 7$^{h}$ & X, opt., radio  \nl
GRB980425 & 8' & 15.1$^{\circ}$ & 0.61 & 40 & 2.4  & 31 & 9$^{h}$ & X, opt, radio ?  \nl
\tablenotetext{(a)}{ Peak fluxes in units of 10$^{-7}$ erg cm$^{-2}$ s$^{-1}$}
\enddata
\end{deluxetable}

\clearpage

\begin{deluxetable}{lcccccc||cccccl}
\small
\footnotesize
\scriptsize
\tablewidth{0pt}
\tablenum{2}
\tablecaption{Spectra of GRBs and associated X--ray afterglow}
\label{table2}
\tablehead{
\# & GRB &
Sect. & 
$-\Gamma_{X}$ \tablenotemark{(a)} & 
$-\Gamma_{\gamma}$ \tablenotemark{(a)} &
E$_{p}$ \tablenotemark{(b)} &
N$_{H}$ \tablenotemark{(b)} &
$-\Gamma_{a}$ &
N$_{H,a}$ &
$\delta$ &
Ref.} 
\startdata
1 & GRB960720  &A & $<$ -0.1 & $<$ -0.1 & $>$ 700 & 0.255 & & & & \nl
 & & B & 0.39 $\pm$ 0.18 & 0.39 $\pm$ 0.18 & $>$ 700  & 0.255 & & & & \nl
 & & C & 0.49 $\pm$ 0.17 & 1.76 $\pm$ 0.32 & 178 $\pm$ 40  & 0.255 & & & & \nl
 & & D & 0.76 $\pm$ 0.21 & 2.21 $\pm$ 0.40 & 28 $\pm$ 17 & 0.255 & & & & \nl
 & & E & 1.18 $\pm$ 0.31 & $>$ 1.64  & $<$ 16.7 & 0.255 & & & & \nl
 & & F & 1.9 $\pm$ 0.6 & $>$ 2.44 & $<$ 3. & 0.255 & \nodata &
\nodata & \nodata &   \nl
 & & & & & & & & & & \nl
2 & GRB970111  &A & $<$ -0.37 & 0.63 $\pm$ 0.33 & $>$ 500 & 0.46 & & & & \nl
 & & B & $<$ -0.4 & 1.18 $\pm$ 0.17 & $>$ 194  & 0.46 & & & & \nl
 & & C & -0.63 $\pm$ 0.36 & 1.23 $\pm$ 0.09 & $>$ 170  & 0.46 & & & & \nl
 & & D & -0.92 $\pm$ 0.31 & 1.72 $\pm$ 0.06 & $>$ 115  & 0.46 & & & & \nl
 & & E & -0.76 $\pm$ 0.39 & 1.73 $\pm$ 0.06 & $>$ 118  & 0.46 & & & & \nl
 & & F & -0.82 $\pm$ 0.27 & 1.89 $\pm$ 0.05 & $>$ 89  & 0.46 & & & & \nl
 & & G & -0.96 $\pm$ 0.29 & 2.38 $\pm$ 0.11 &  37 $\pm$ 12 & 0.46 & & & & \nl
 & & H & -0.87 $\pm$ 0.27 & 2.77 $\pm$ 0.11 & 33 $\pm$  8  & 0.46 & & & & \nl
 & & I & -0.19 $\pm$ 0.17 & 2.9 $\pm$ 0.2 & 36 $\pm$ 9 & 0.46 & & & & \nl
 & & J & 0.33 $\pm$ 0.12 & 3.05 $\pm$ 0.27 & 35 $\pm$ 7 & 0.46 & \nodata &
\nodata & $>$ 1.5 & (1)  \nl
 & & & & & & & & & & \nl
3 & GRB970228  &A & 0.92 $\pm$ 0.03 & 1.54 $\pm$ 0.18 & $>$ 700 & 1.6 & & & & \nl 
 & & B & 1.4 $\pm$ 0.1 & 2.5 $\pm$ 0.1 & 35 $\pm$ 18 & 1.6 & & & & \nl
 & & C & 1.8 $\pm$ 0.1 & 1.8 $\pm$ 0.1 & $<$ 2 & 1.6 & & & & \nl
 & & D & 1.84 $\pm$ 0.09 & 1.84 $\pm$ 0.09 & $<$ 2 & 1.6 & & & & \nl
 & & E & 1.92 $\pm$ 0.15 & 1.92 $\pm$ 0.15 & $<$ 2 & 1.6 & & & & \nl
 & & F & 1.5 $\pm$ 0.4 & $>$ 0.6 & $<$ 2 & 1.6 & & & & \nl
 & & G & 1.6 $\pm$ 0.1 & 1.4 $\pm$ 0.3 & $<$ 2 & 1.6 & 2.06 $
\pm$ 0.24 & 3.5 $\pm$ 2.3 & 1.33 $\pm$ 0.12 & (2) \nl 
 & & & & & & & & & & \nl
4 & GRB970402 &  A & 1.38 $\pm$ 0.03 & 1.38 $\pm$ 0.03 & $\geq$700 & $<$ 20 \nl
 & & B & 1.36 $\pm$ 0.04 & 1.36 $\pm$ 0.04 & $\geq$700 & 2.1 & 1.7 $\pm$
 0.6 & 2.1 & 1.56 $\pm$ 0.03 & (3) \nl
 & & & & & & & & & & \nl
5 & GRB970508 &  A & 0.83 $\pm$ 0.11 & 0.83 $\pm$ 0.11 & $>$ 700 & 0.51 & & & & \nl
 & & B & 1.54 $\pm$ 0.10 & 1.54 $\pm$ 0.10 & $\geq$700 & 0.51 & & & & \nl
 & & C & 1.74 $\pm$ 0.14 & 1.74 $\pm$ 0.14 & $\geq$700 & 0.51 & & & & \nl
 & & D & 1.8 $\pm$ 0.4 & $>$ 1.4 & $<$ 24 & 0.51 & 2.1 
$\pm$ 0.6 & 10. $\pm$ 5. & 1.1 $\pm$ 0.1 & (4),(5) \nl
 & & & & & & & & & & \nl
6 & GRB971214 &  A & 0.37 $\pm$ 0.23 & 1.4 $\pm$ 0.03 & $>$ 700 & 0.16 & & & & \nl
 & & B & 0.33 $\pm$ 0.27 & 1.0 $\pm$ 0.07 & $>$ 224 & 0.16 & & & & \nl
 & & C & 0.96 $\pm$ 0.09 & 2.6 $\pm$ 0.5 & 56 $\pm$ 10 & 0.16 & 1.7 $\pm$ 0.
2 & 0.9 $\pm$ 0.5 & 1.2 $\pm$ 0.13 & (6),(7) \nl
 & & & & & & & & & & \nl
7 & GRB980329 &  A & $<$ 0.16 & 1.32 $\pm$ 0.03 & $>$ 229 & 120 
$\pm$ 80 & & & & \nl
 & & B & 0. $\pm$ 0.33 & 1.30 $\pm$ 0.06 & $>$ 168 & 120 $\pm$ 80 & & & & \nl
 & & C & 0.36 $\pm$ 0.19 & 1.25 $\pm$ 0.08 & $>$ 210 & 120 $\pm$ 80 & & & & \nl
 & & D & 0.67 $\pm$ 0.11 & 1.29 $\pm$ 0.02 & $>$ 320 & 120 $\pm$ 80 & & & & \nl
 & & E & 0.73 $\pm$ 0.18 & 1.61 $\pm$ 0.04 & $>$ 213 & 250 $\pm$ 100  & & & & \nl
 & & F & 0.78 $\pm$ 0.32 & 1.70 $\pm$ 0.09 & $>$ 180 & 85 $\pm$ 40  & & & & \nl
 & & G & 1.0 $\pm$ 0.4 & 1.7 $\pm$ 0.2 & $>$ 175 & 10.  & & & & \nl
 & & H & 1.5 $\pm$ 0.5 & 2.3 $\pm$ 0.3 & 105 $\pm$ 80. & 10. & 2.4 $\pm$ 0.4 & 10 $\pm$ 0.4 & 1.35 $\pm$ 0.03 & (8) \nl 
 & & & & & & & & & & \nl
8 & GRB980425 &  A & $<$ 0.3 & 1.75 $\pm$ 0.15 & $>$ 68 & 0.396 & & & & \nl
 & & B & 0.78 $\pm$ 0.27 & 2.3 $\pm$ 0.1 & 68 $\pm$ 40 & 0.396 & & & &  \nl
 & & C & 1.45 $\pm$ 0.75 & 3.3 $\pm$ 0.7 & 31 $\pm$ 25 & 0.396 & & & &  \nl
 & & D & 1.6 $\pm$ 0.5 & $>$ 3.8 & $<$ 57 & 0.396 & \nodata & \nodata & $>$1.3 & (9) \nl
\enddata
\tablenotetext{}{$^{1}$Feroci et al.~1998 \nocite{Feroci98}, 
$^{2}$Frontera et al.~1998 \nocite{Frontera98a},
 $^{3}$Nicastro et al.~ 1998 \nocite{Nicastro98},
 $^{4}$Piro et al.~1998 \nocite{Piro98b},
 $^{5}$Amati et al. 1998 \nocite{Amati98},
 $^{6}$Heise et al. 1998 \nocite{Heise98},
 $^{7}$Dal Fiume et al.~1999\nocite{Dalfiume99},
 $^{8}$in 't Zand et al.~1998 \nocite{Zand98},
 $^{9}$Pian et al.~1999a \nocite{Pian99a}
}
\tablenotetext{a}{The photon indices reported in columns 2 and 3 refer to fit
with the smoothed broken power-law proposed by Band et al. (1993)} 
\tablenotetext{b}{$E_{p}$ values are in keV; $N_{H}$ values are in units of
 $10^{21}\, cm^{-2}$}
\tablenotetext{c}{Data referring to the first impulse of the event (see text 
and paper from Frontera et al.~1998a)}
\end{deluxetable}

\clearpage

\begin{deluxetable}{cccccccccc}
\small
\footnotesize
\scriptsize
\tablewidth{0pt}
\tablenum{3}
\tablecaption{GRB and X--ray afterglow energetics \tablenotemark{(a)}}
\label{table3}
\tablehead{
GRB & S$_{X}$ & S$_{\gamma}$ & S$_{a}$ &S$_{X}$/S$_{\gamma}$ &  S$_{a}$/S$_{\gamma}$ 
& S$_{a}$/S$_{X}$ & E$_{52}$\tablenotemark{(b)} & z & Ref.\tablenotemark{(c)} \nl
 & (2--10 keV) & (40--700 keV) & (2--10 keV) & (\%) & (\%) & & & &  } 
\startdata
GRB960720 & 0.08 $\pm$ 0.02 & 2.6 $\pm$ 0.3 & & 3.0 $\pm$ 0.8  & \nodata  & \nodata  & \nodata  & \nodata  & \nodata  \nl
GRB970111 & 1.6 $\pm$ 0.1 & 43 $\pm$ 3. & $<$ 1.1 &3.7 $\pm$ 0.2 &  $<$ 3 & $<$ 0.69 & \nodata  & \nodata  & \nodata  \nl
GRB970228 & 2.2 $\pm$ 0.5 & 11 $\pm$ 1. & 2.16 $\pm$ 0.81 & 20 $\pm$ 0.81 & 19.6 $\pm$ 0.2 & 0.98 $\pm$ 0.19 & 0.25 & 0.695 & (1)\nl
GRB970402 & 0.4 $\pm$ 0.04 & 8.2 $\pm$ 0.9 & 0.47 $\pm$ 0.15 & 4.8 $\pm$ 0.1 &  5.7 $\pm$ 0.7 & 1.17 $\pm$ 0.13 & & & \nl
GRB970508 & 0.33 $\pm$ 0.05 & 1.8 $\pm$ 0.3 & 0.47 $\pm$ 0.31 & 38.9 $\pm$ 1.8 &  26 $\pm$ 12 & 0.67 $\pm$ 0.30 & $\sim$0.078 & 0.835 & (2) \nl
GRB971214 & 0.19 $\pm$ 0.03 & 8.8 $\pm$ 0.9 & 0.35 $\pm$ 0.14 & 1.2 $\pm$ 0.1 &  3.9 $\pm$ 0.7 & 3.20 $\pm$ 0.75 &  $\sim$0.936 & 3.418 & (3) \nl
GRB980329 & 0.7 $\pm$ 0.12 & 55. $\pm$ 5. & 0.72 $\pm$ 0.27 & 1.0 $\pm$ 0.04 &  0.40 $\pm$ 0.07 & 0.39 $\pm$ 0.8 &  $\sim$3.86 (?)  & $\sim$ 5 (?) & (4) \nl
GRB980425 & 0.78 $\pm$ 0.02 & 2.8. $\pm$ 0.5 & $<$0.46 & 28 $\pm$ 5 &  $<$16 & $<$0.59 &  $<$ 7.45x10$^{-6}$ & 0.008 & (5) \nl
\enddata
\tablenotetext{a}{All fluences are given in units of 10$^{-6}$ erg/cm$^{2}$}
\tablenotetext{b}{E$_{52}$ = afterglow energy in units of 10$^{52}$ erg}
\tablenotetext{c}
{References for redshift: $^{1}$Djorgovski et al., 1999 \nocite{Djorgovski99},
$^{2}$Metzger et al., 1997 \nocite{Metzger97},
 $^{3}$Kulkarni et al. 1998a \nocite{Kulkarni98a},
 $^{4}$Fruchter, 1999 \nocite{Fruchter99},
 $^{5}$Galama et al., 1998b} \nocite{Galama98b}
\end{deluxetable}

\clearpage

\begin{table}
\vspace{-2cm}
\epsscale{1.0}
\plotone{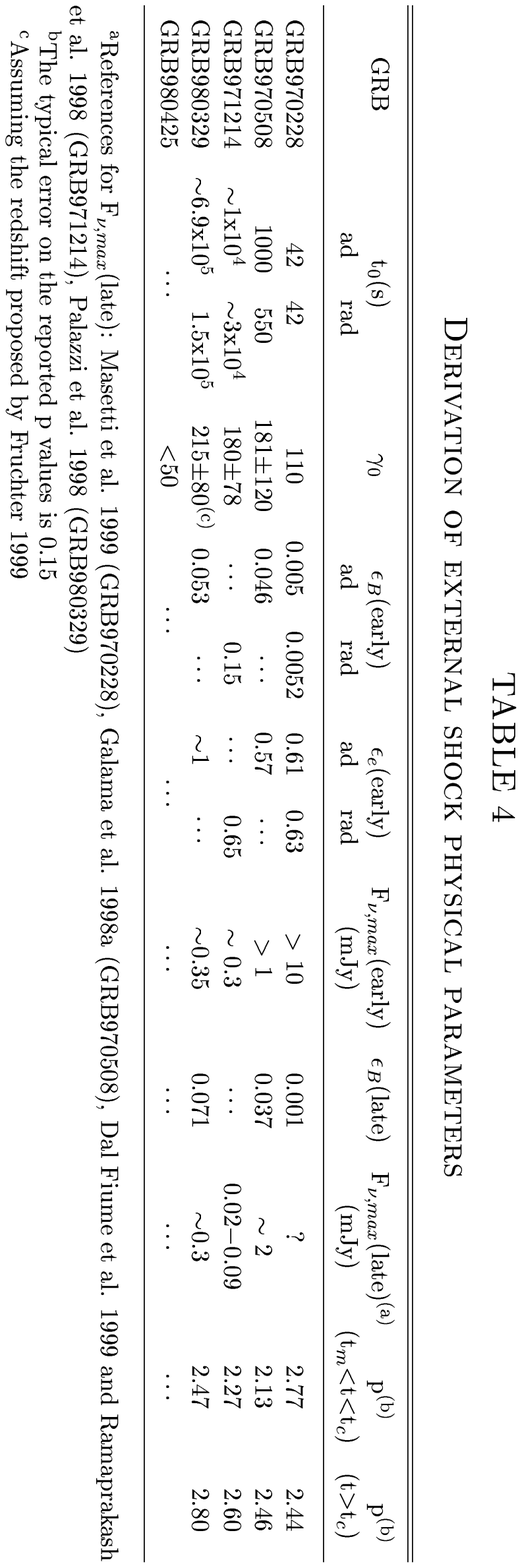}
\ifPreprintMode
\caption[]{}
\label{tab4}
\fi
\end{table}

\clearpage

\begin{figure}
\figurenum{1}
\epsscale{0.80}
\plotone{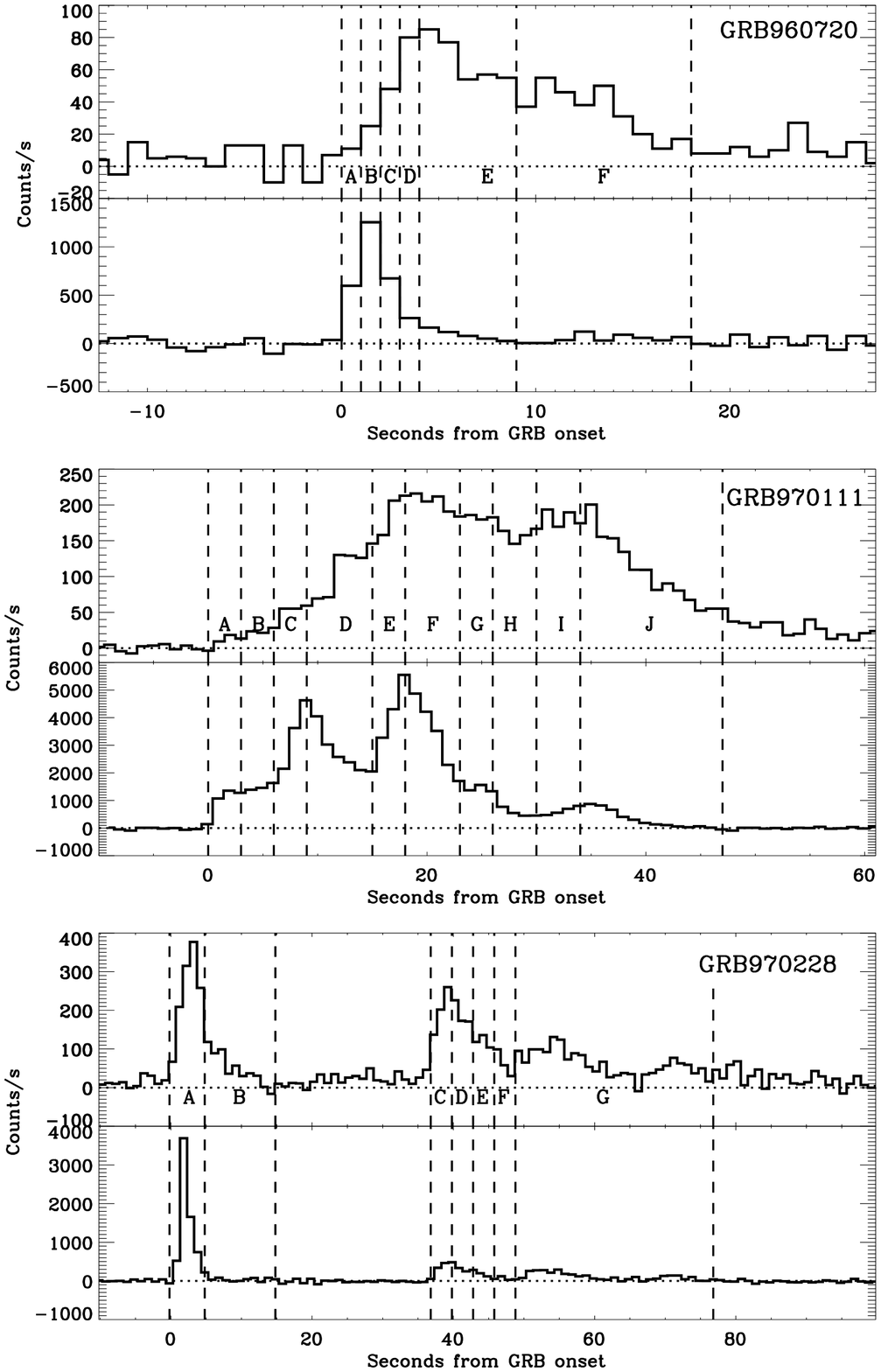}
\caption[fig1a.ps,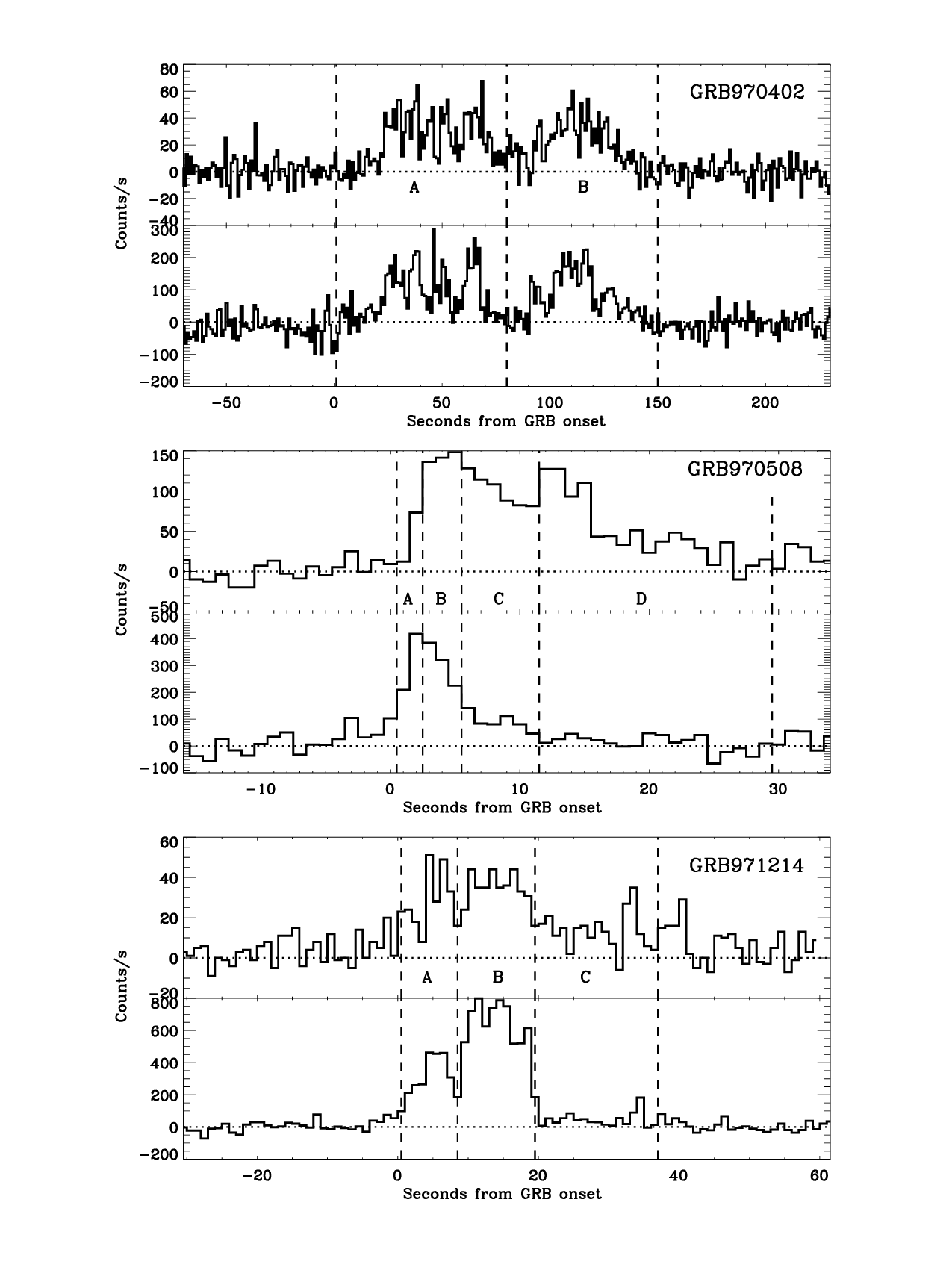,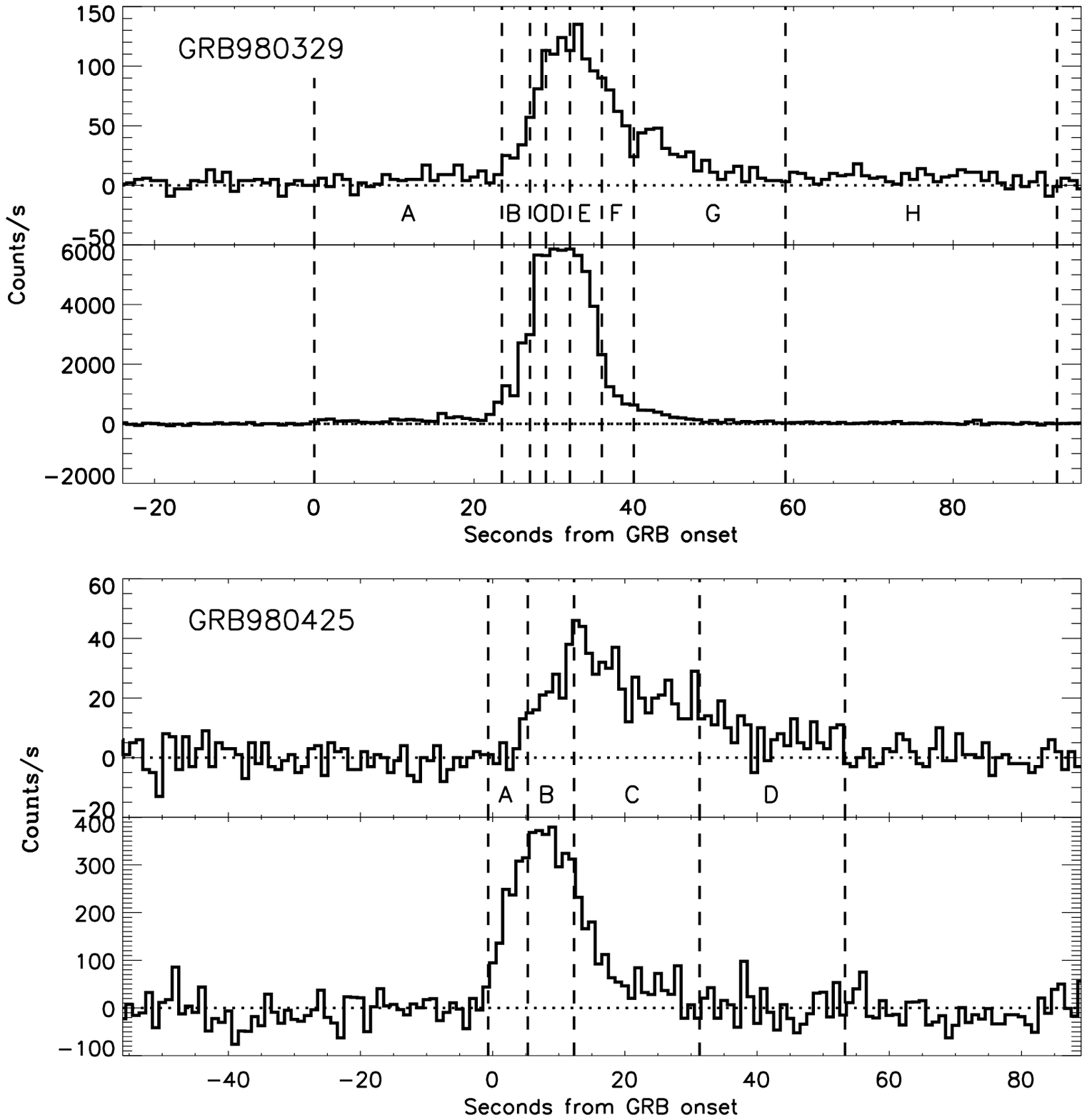]{Light curves of GRBs in our sample,
after background subtraction, in the 2--26 keV (upper panels) 
and 40--700 keV (lower panels) energy bands. Horizontal dotted line gives the 
background reference of each GRB. Vertical dashed lines indicate the 
temporal sections in which we 
performed the spectral analysis.
}
\label{fig1a}
\end{figure}

\clearpage

\begin{figure}
\figurenum{1--continued}
\epsscale{0.80}
\plotone{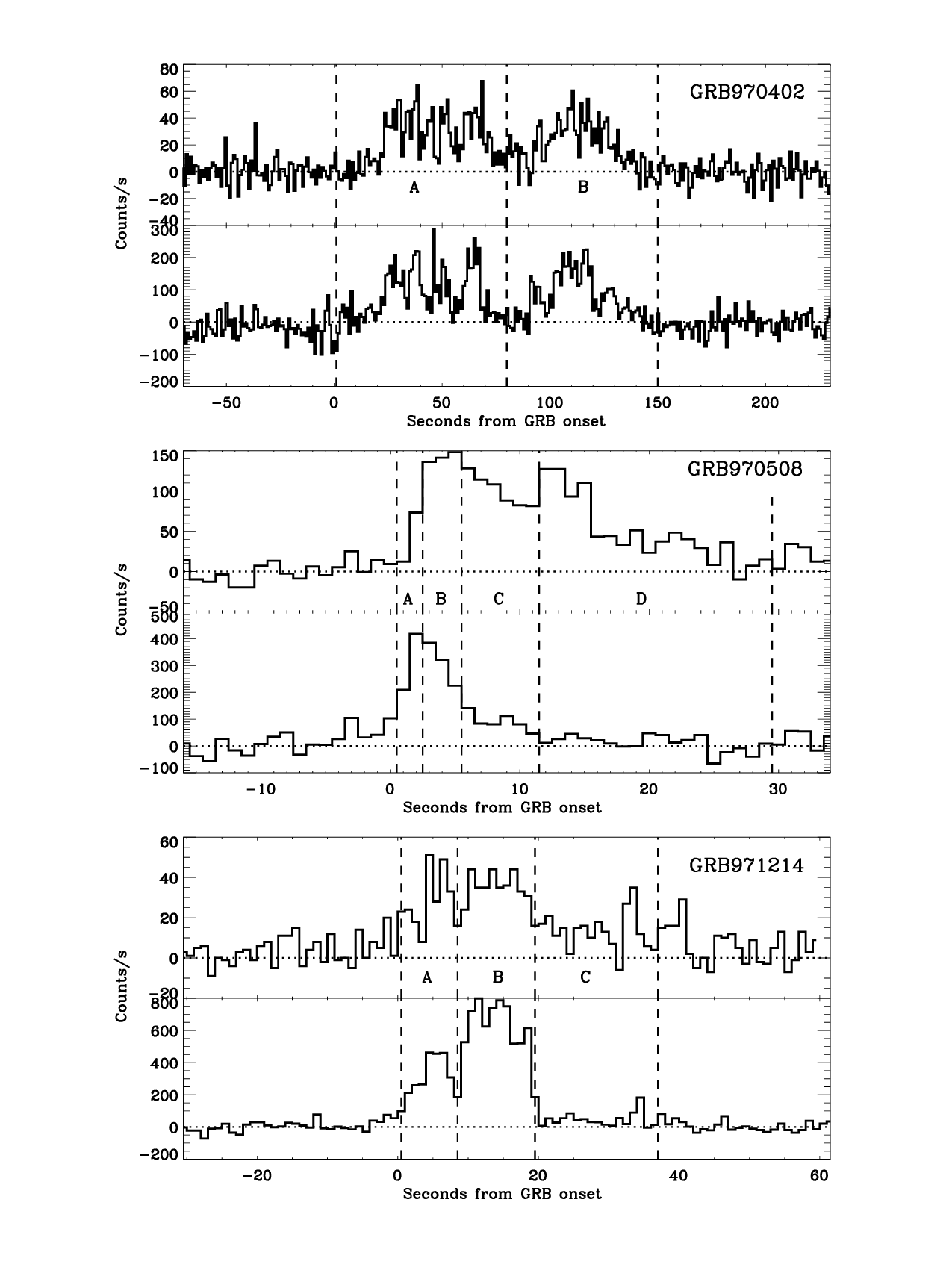}
\caption[]{}
\vspace{1.2cm}
\label{fig1--continued}
\end{figure}

\clearpage

\begin{figure}
\figurenum{1--continued}
\epsscale{0.80}
\plotone{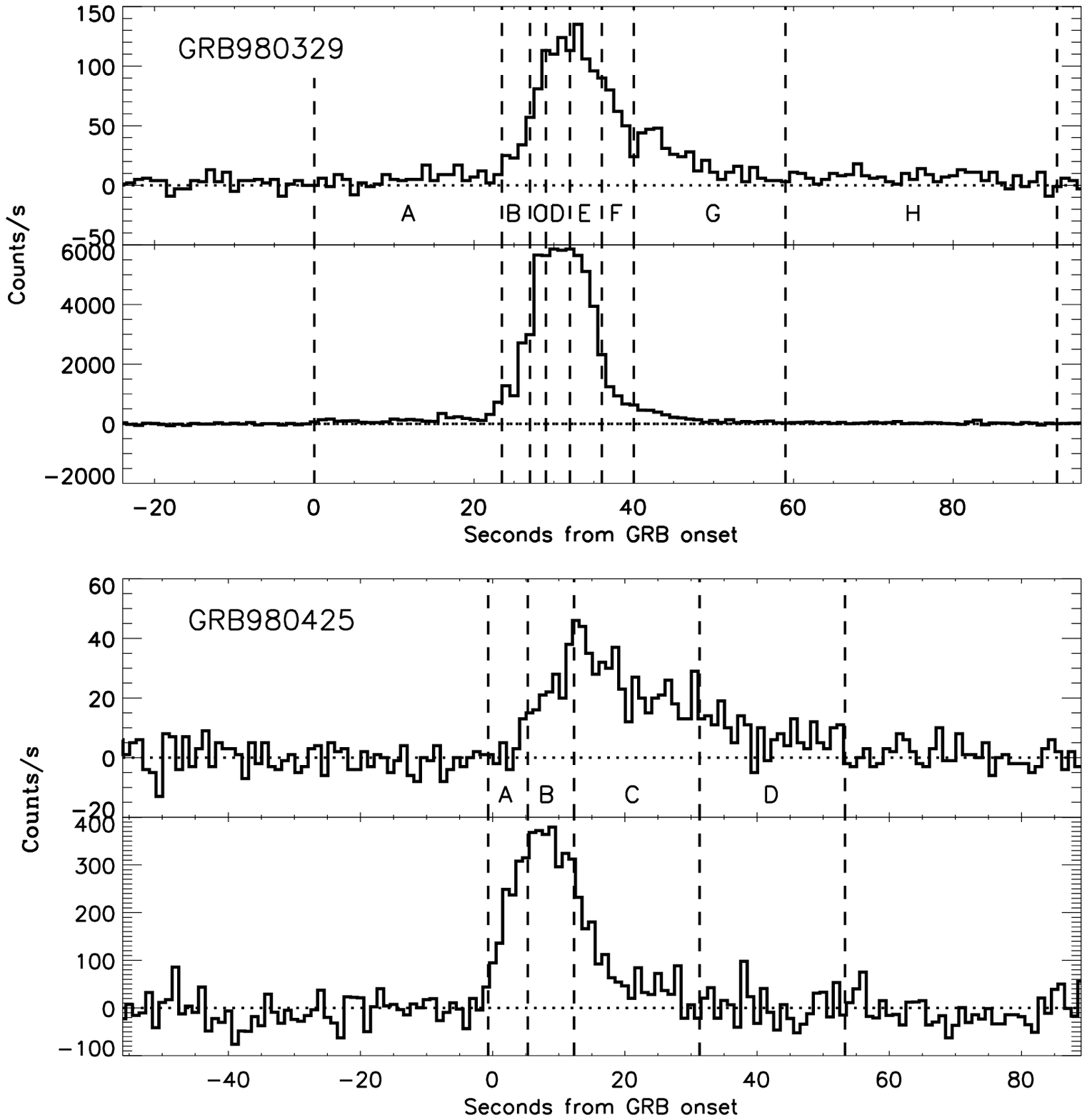}
\caption[]{}
\vspace{1.2cm}
\label{fig1--continued}
\end{figure}

\clearpage

\begin{figure}
\figurenum{2}
\epsscale{0.40}
\plotone{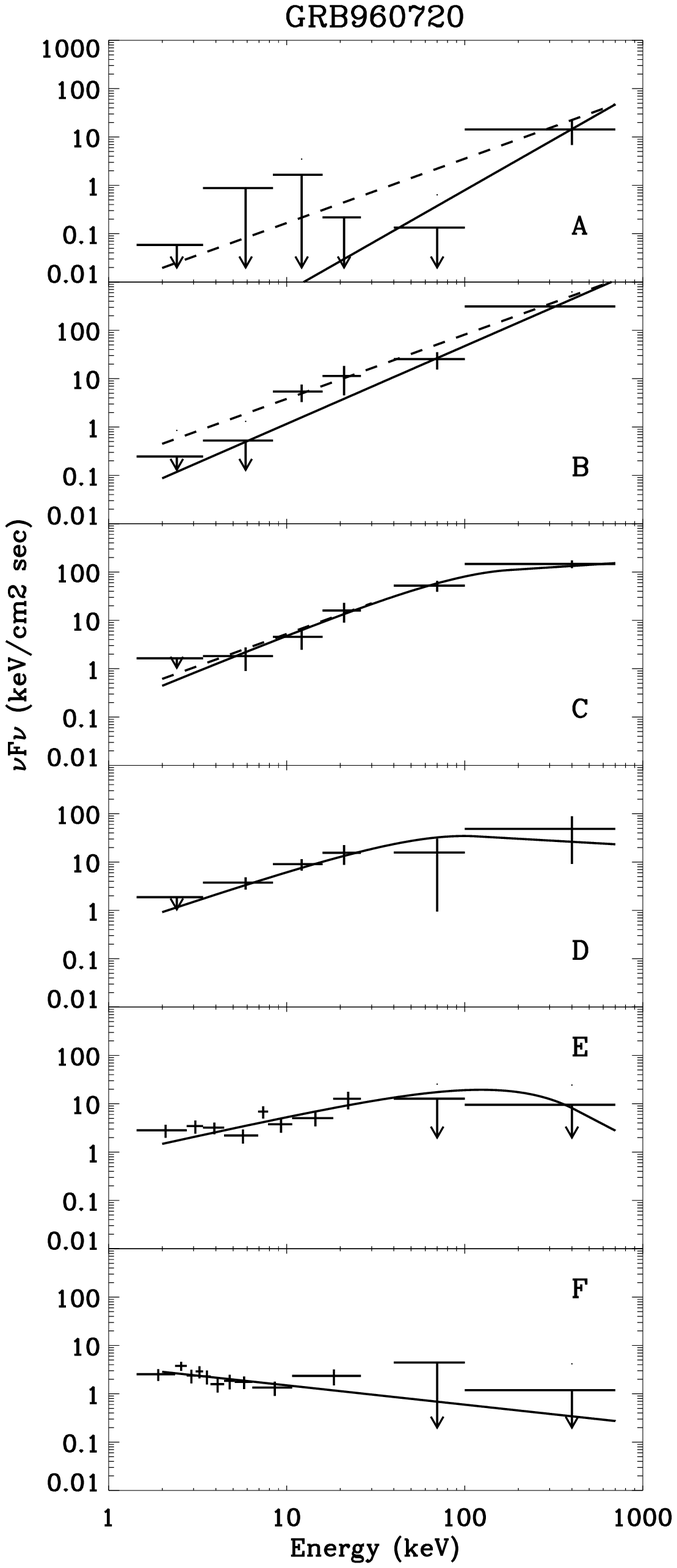}
\vspace{1.0cm}
\caption[fig2a.ps,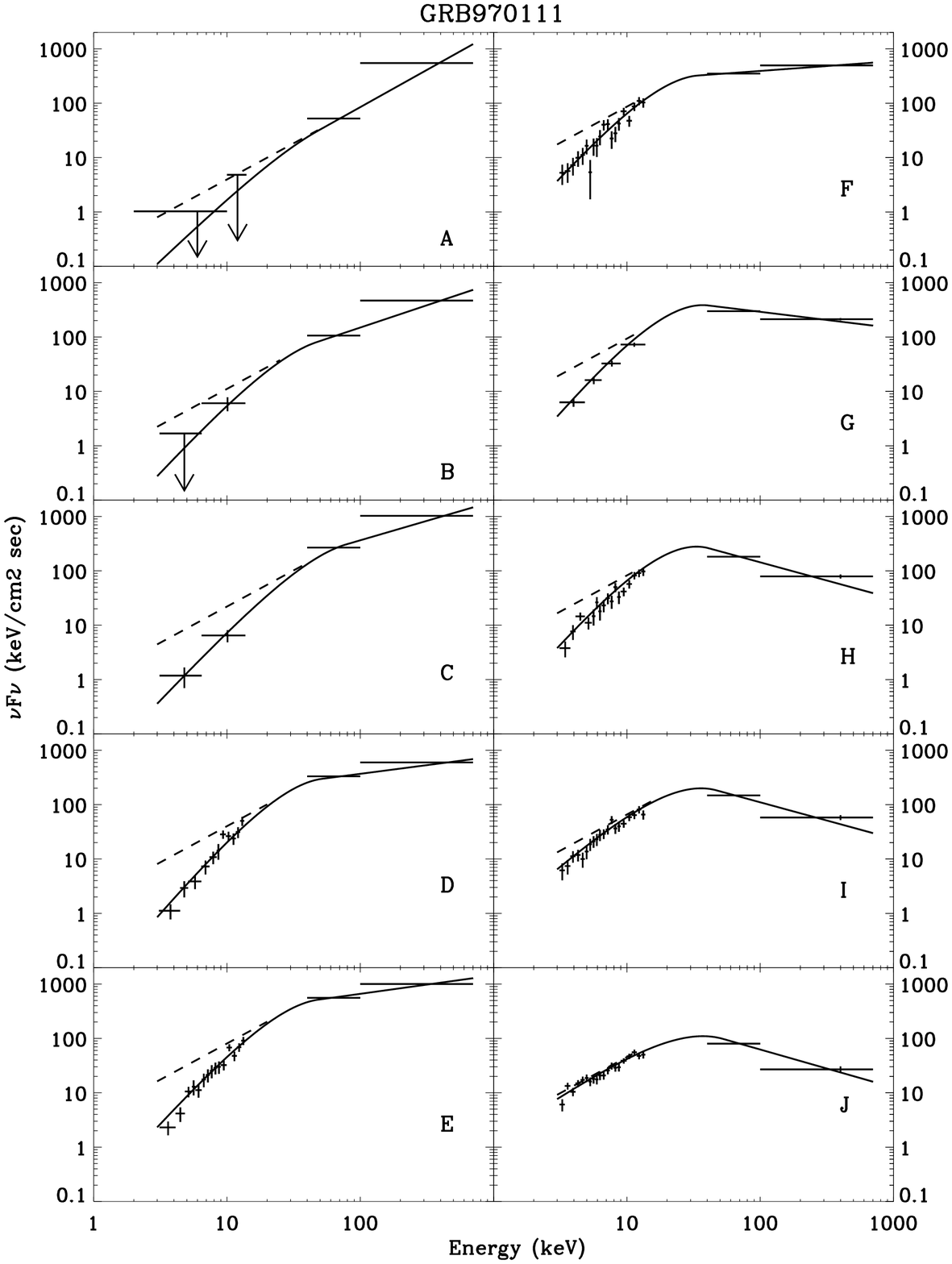,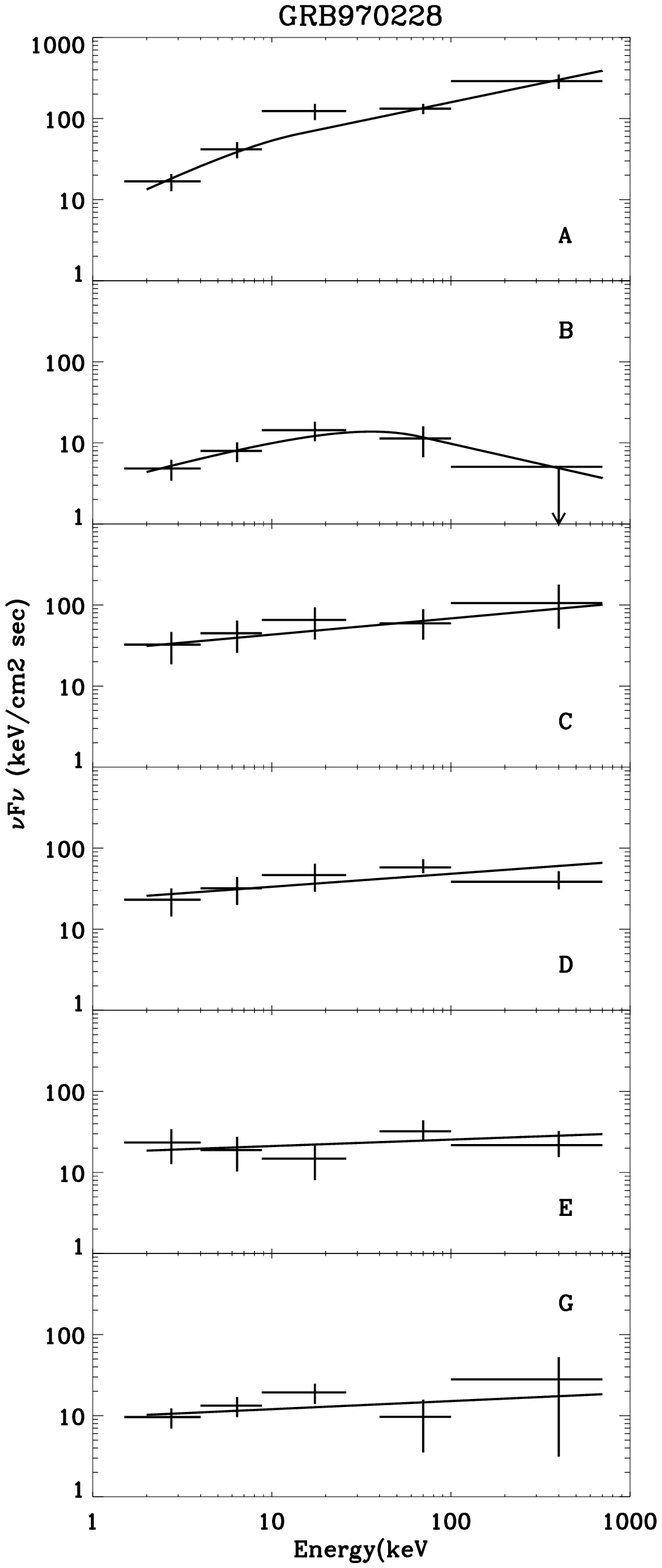,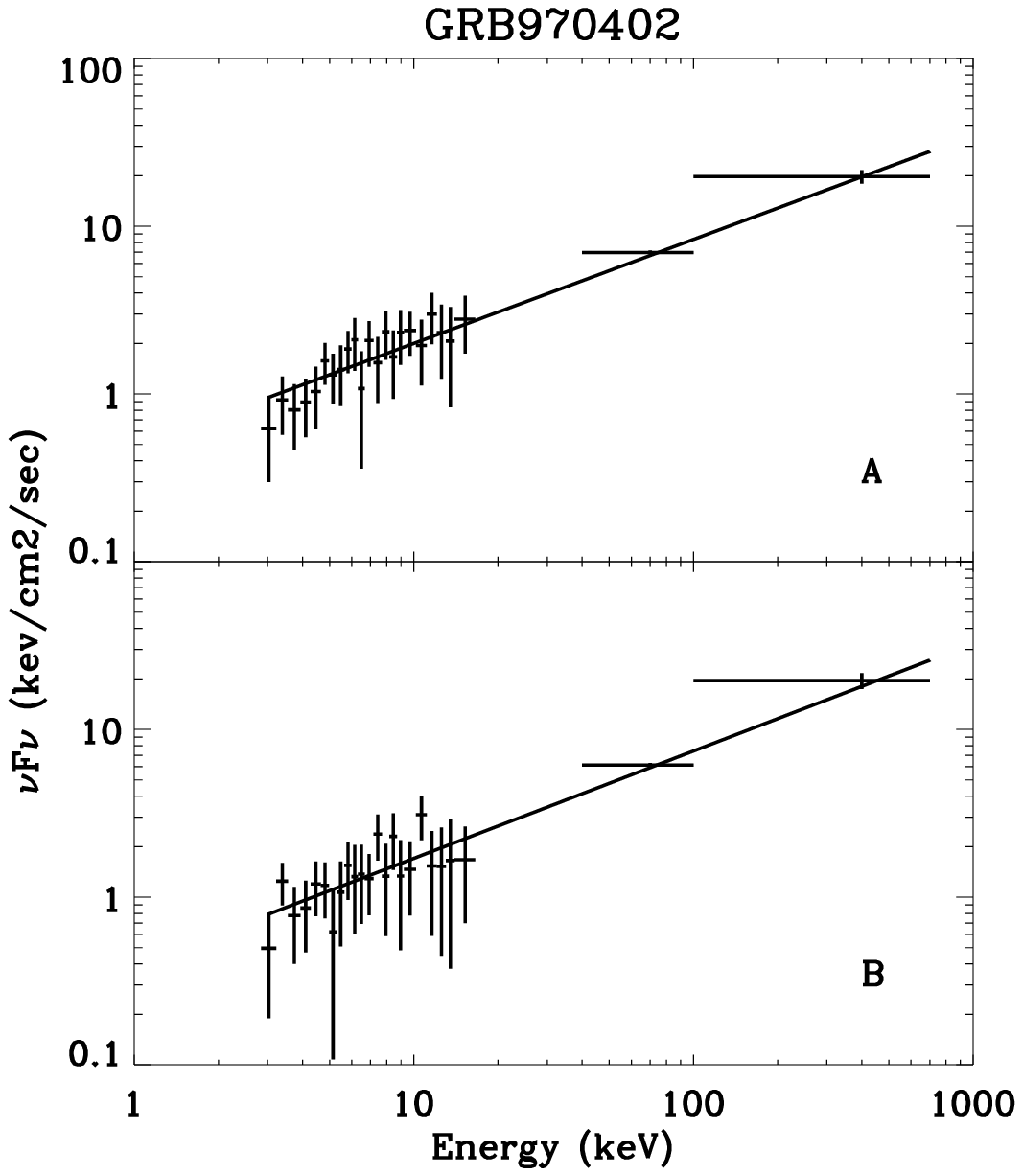,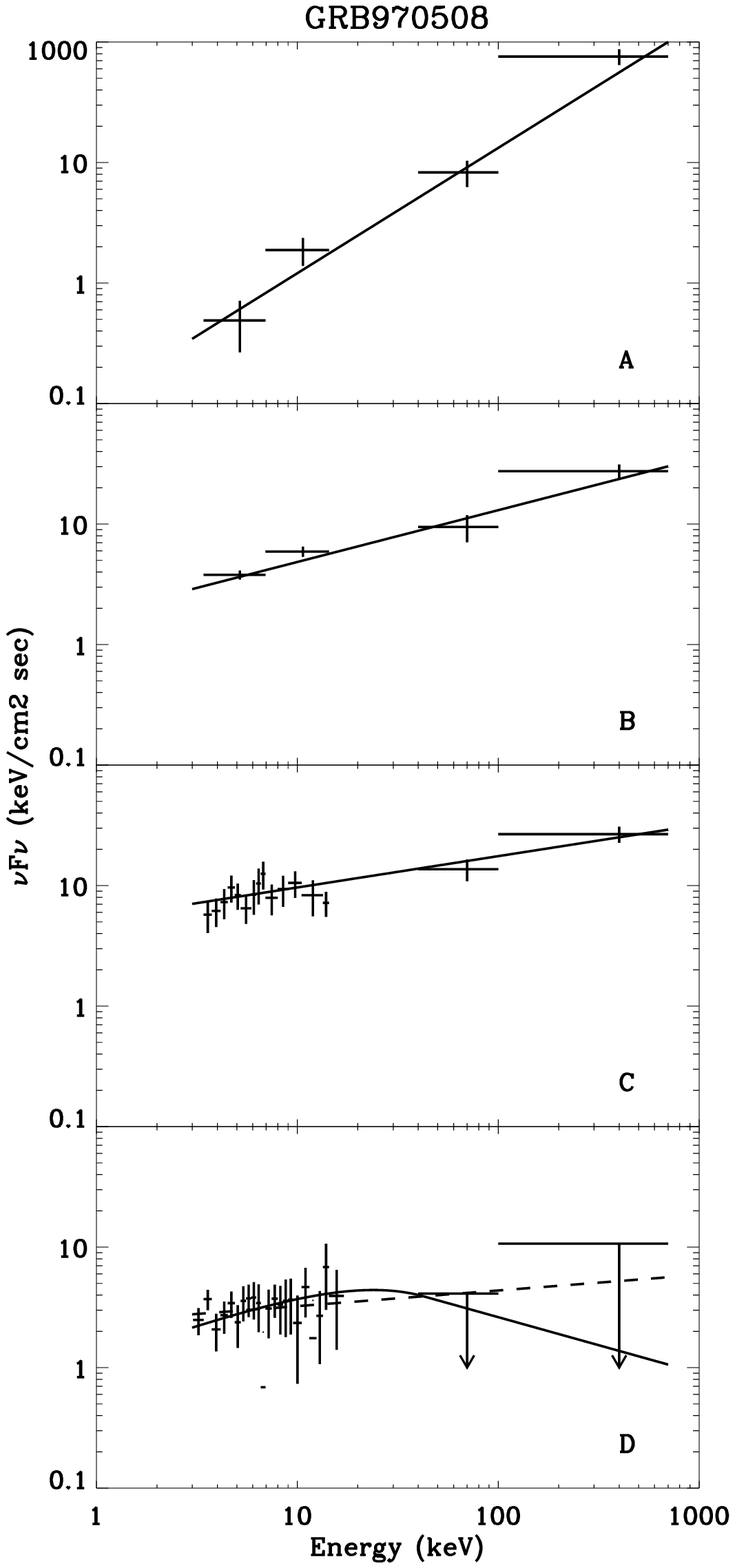,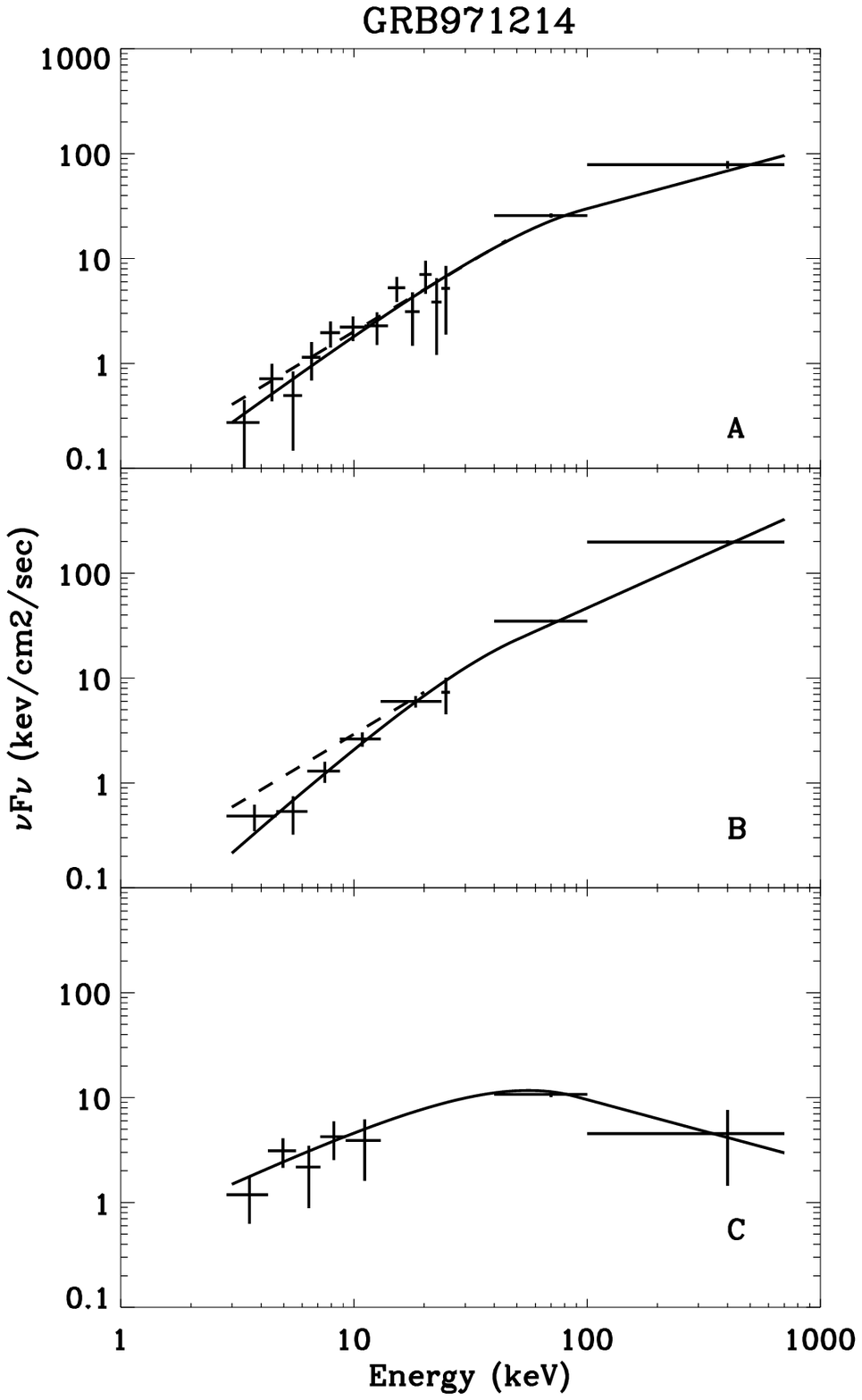,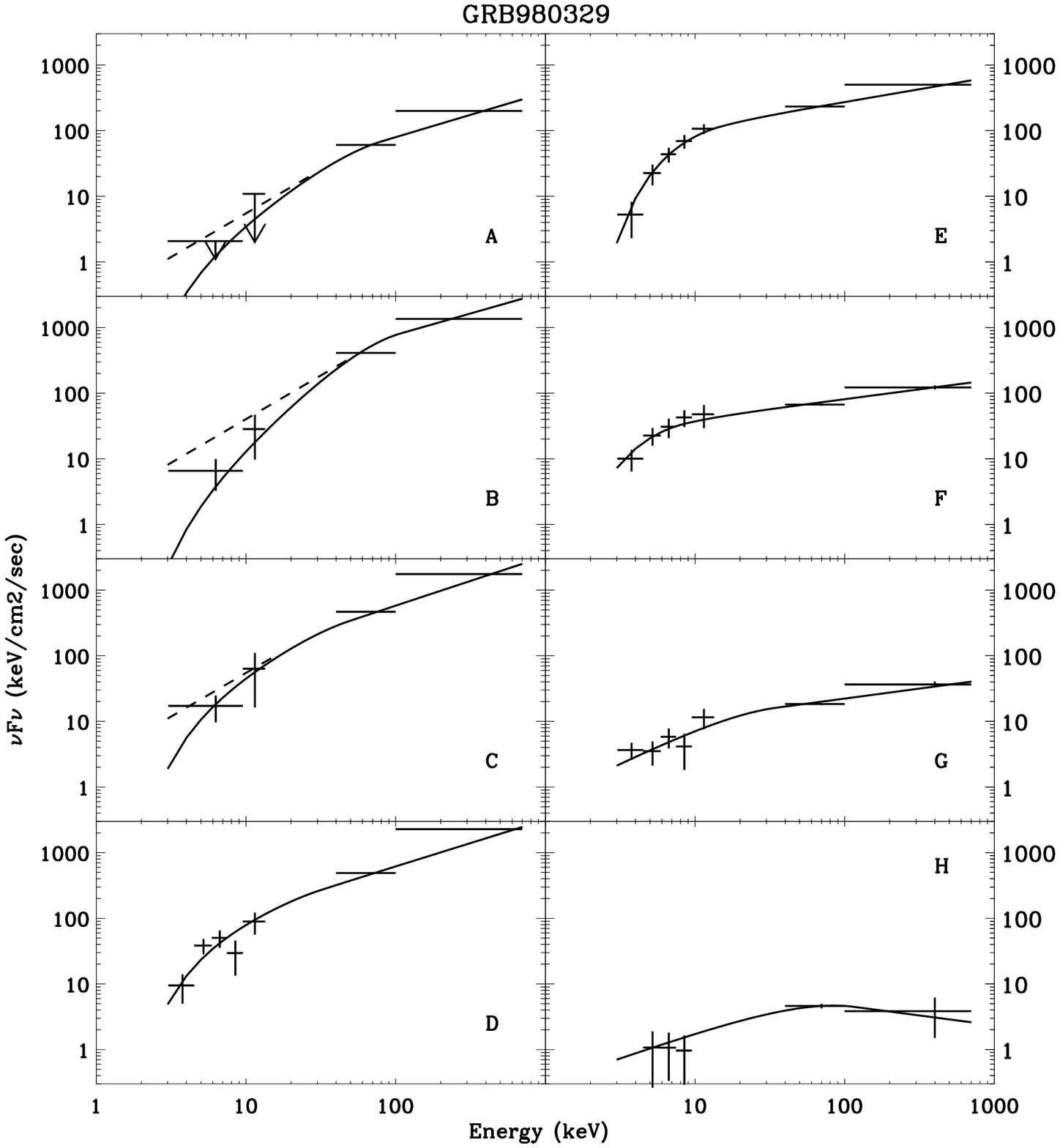,
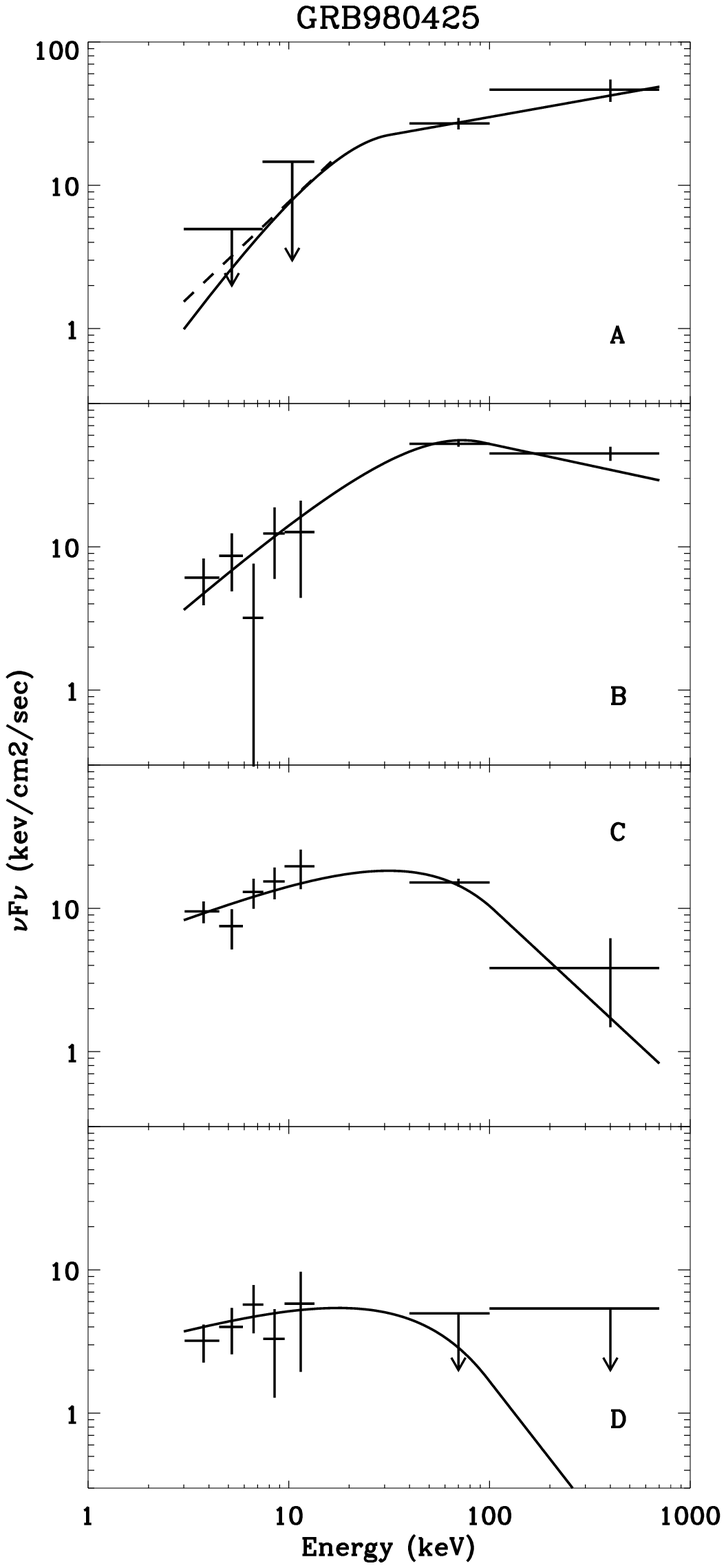]{Spectral energy distributions ($\nu$F($\nu$)) of GRBs in the
2--700~keV. For each GRB, in the different panels are reported
the spectra in the temporal sections in which we divided the GRB light curve. 
Solid lines give
the best fit of the Band law \cite{Band93} to the data. Dashed lines show 
the limit slope in the case of an optically thin synchrotron model (see text).
}
\label{fig2}
\end{figure}

\clearpage

\begin{figure}
\figurenum{2--continued}
\epsscale{0.75}
\plotone{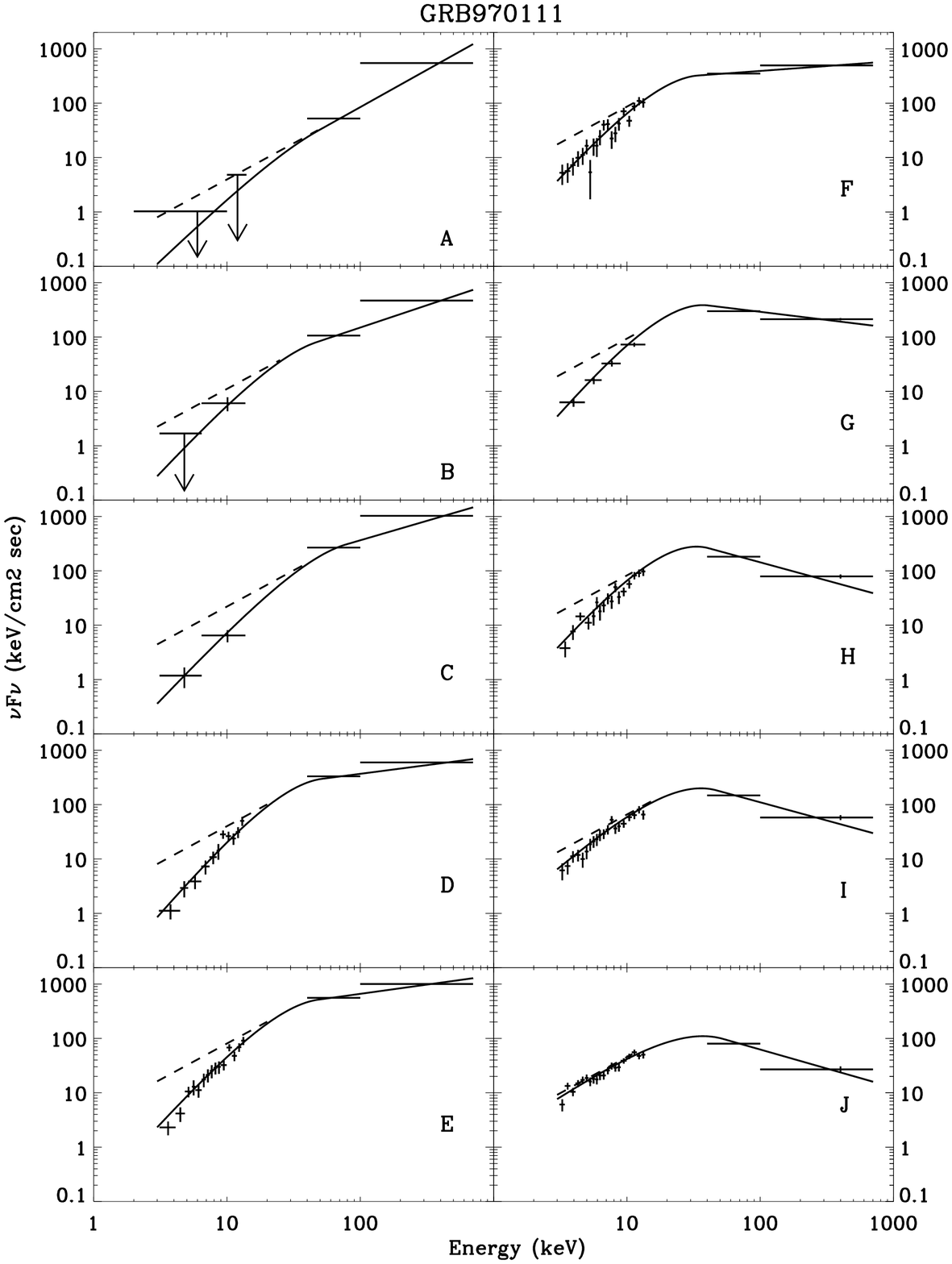}
\vspace{0.5cm}
\caption[]{}
\vspace{1.5cm}
\label{fig2--continued}
\end{figure}

\clearpage

\begin{figure}
\figurenum{2--continued}
\epsscale{0.40}
\plotone{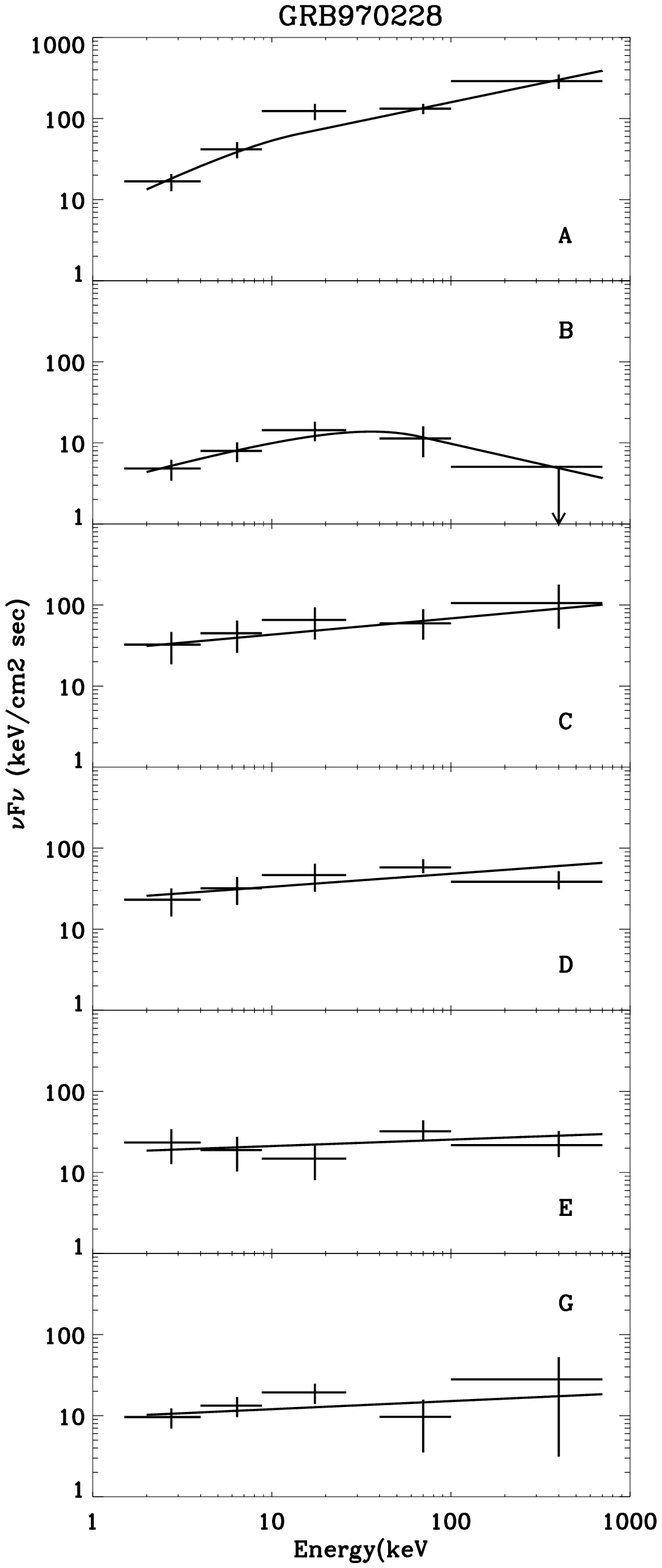}
\caption[]{}
\vspace{1.2cm}
\label{fig2--continued}
\end{figure}

\clearpage

\begin{figure}
\figurenum{2--continued}
\epsscale{0.40}
\vspace{0.7cm}
\plotone{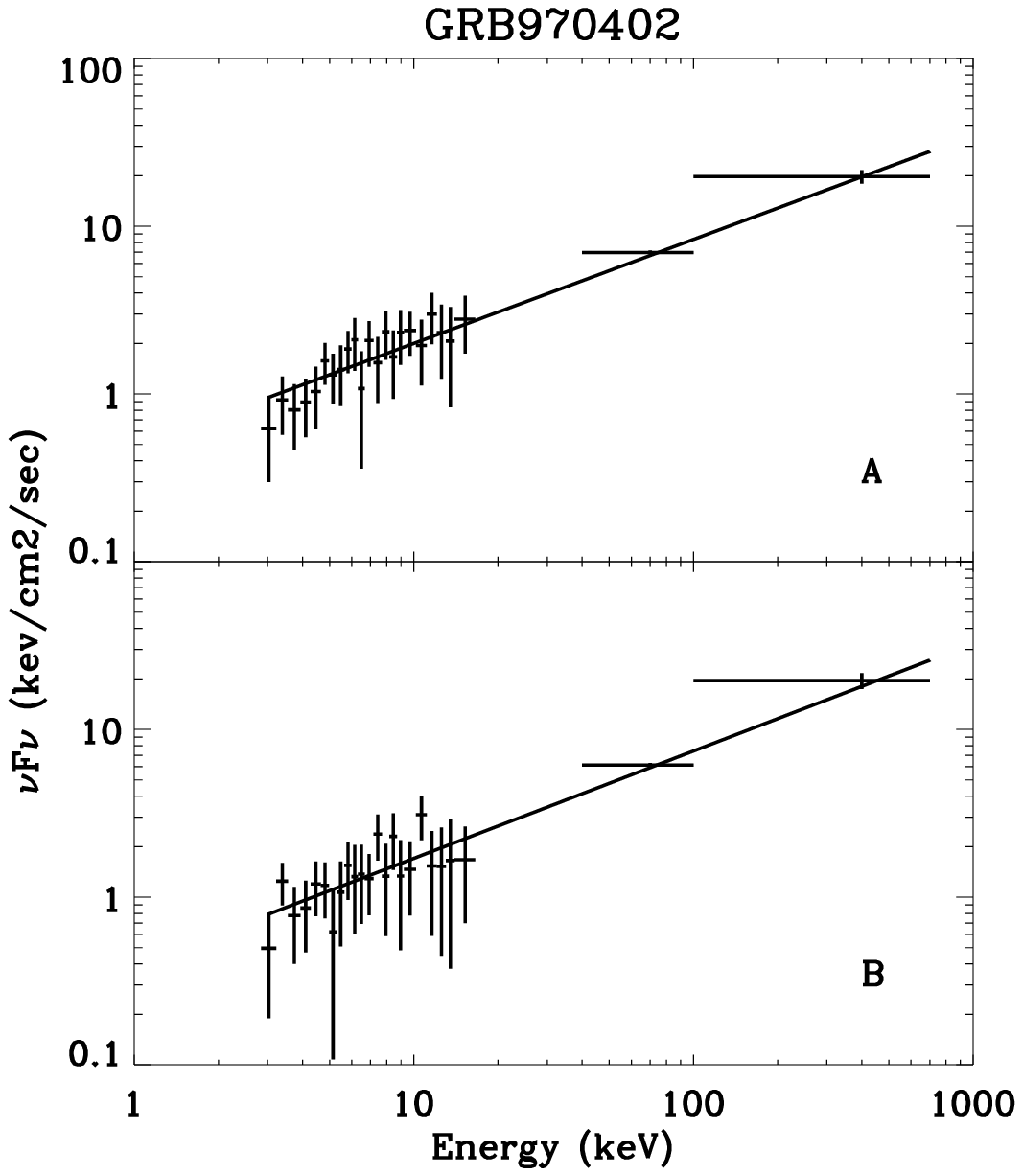}
\caption[]{}
\vspace{1.2cm}
\label{fig2--continued}
\end{figure}

\clearpage

\begin{figure}
\figurenum{2--continued}
\epsscale{0.40}
\vspace{0.7cm}
\plotone{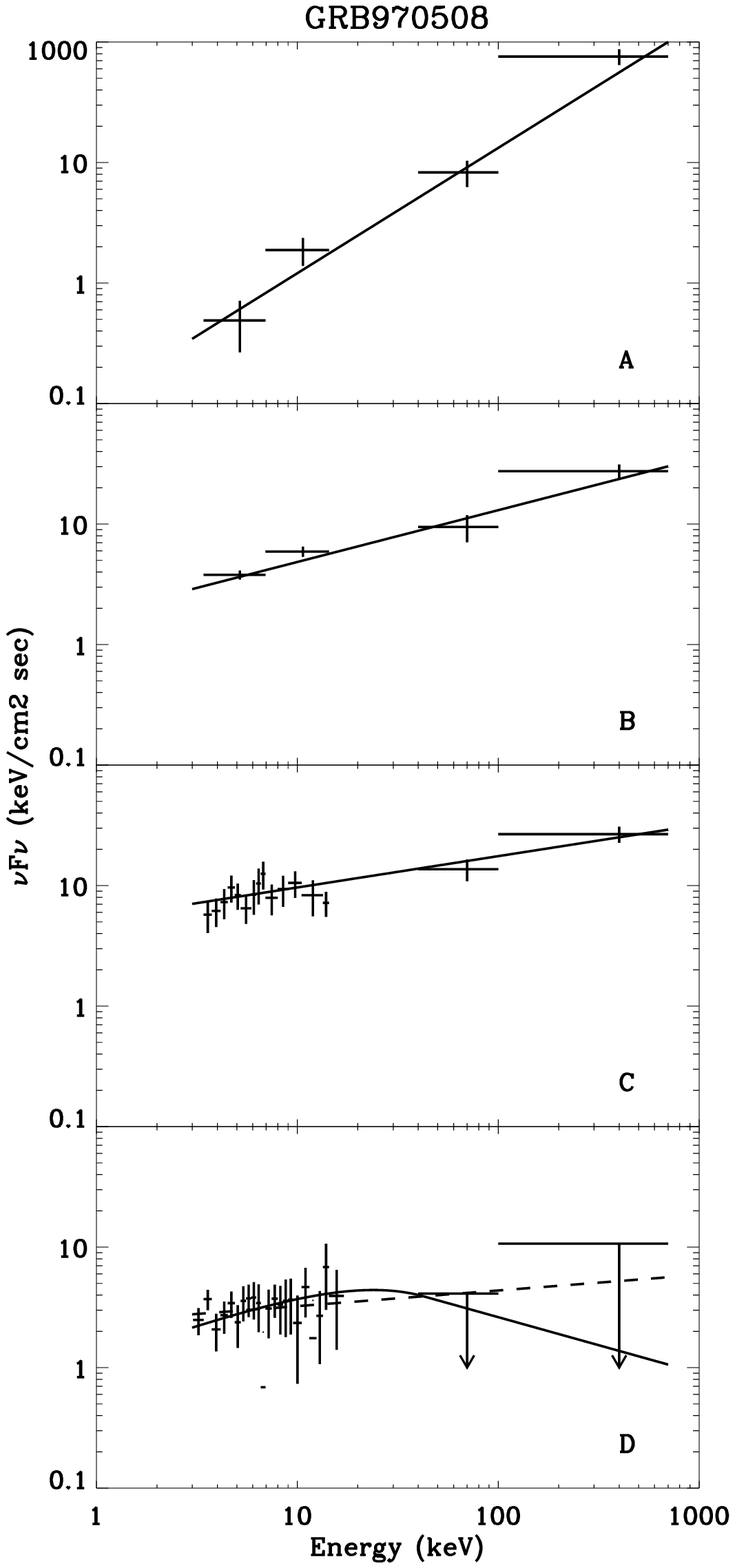}
\caption[]{}
\vspace{1.2cm}
\label{fig2--continued}
\end{figure}

\clearpage

\begin{figure}
\figurenum{2--continued}
\epsscale{0.40}
\vspace{0.7cm}
\plotone{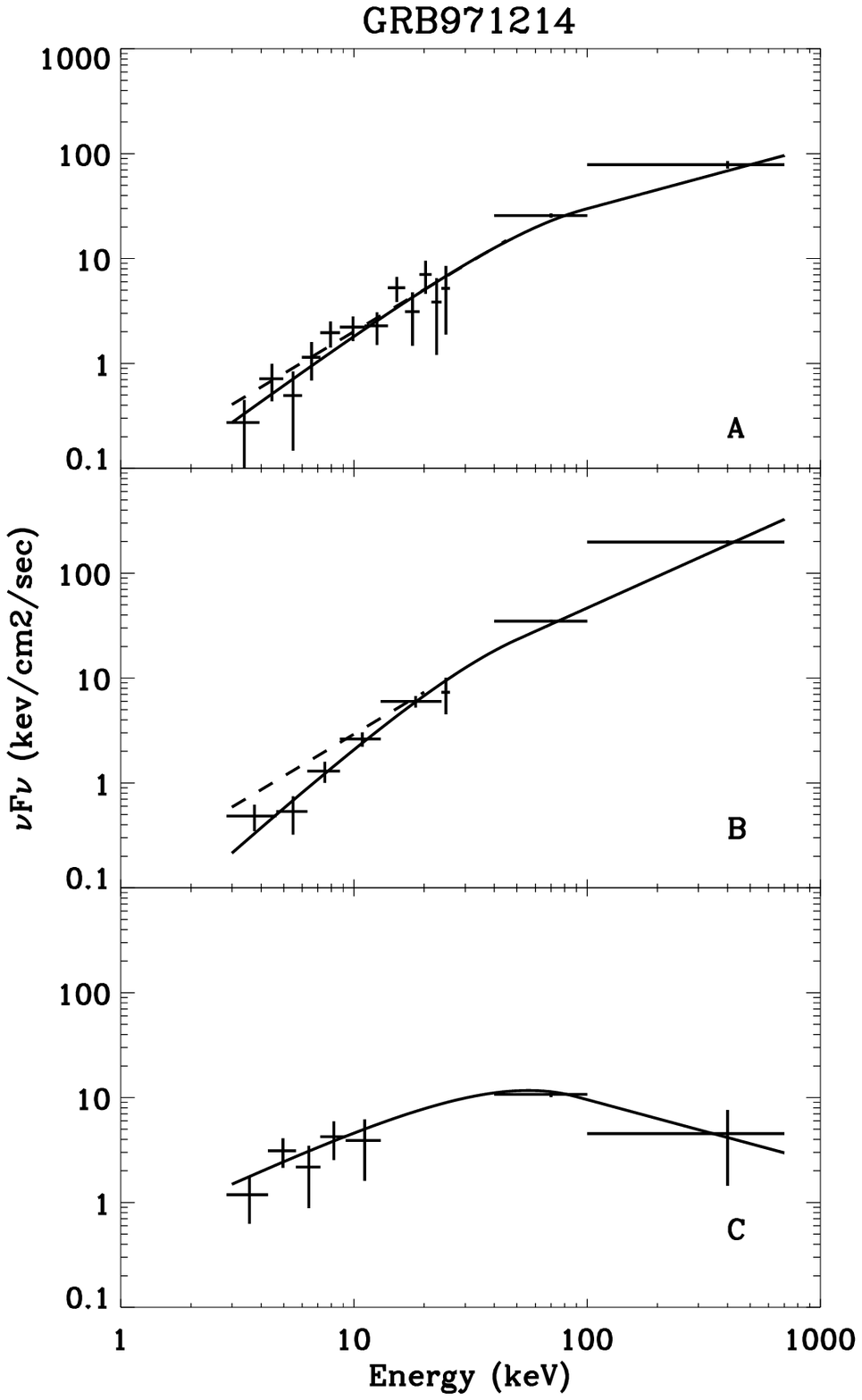}
\caption[]{}
\vspace{1.2cm}
\label{fig2--continued}
\end{figure}

\clearpage

\begin{figure}
\figurenum{2--continued}
\epsscale{0.75}
\plotone{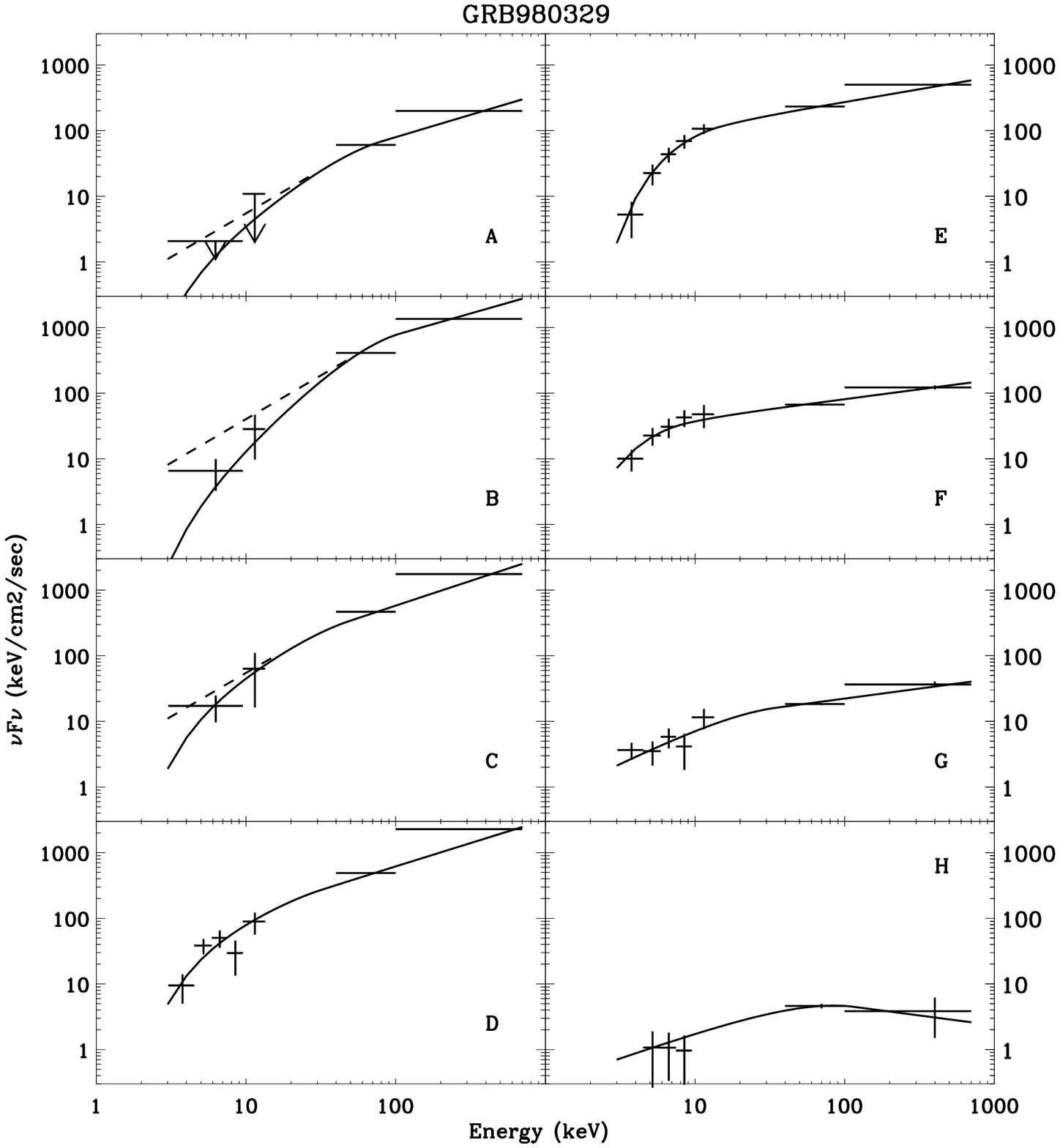}
\caption[]{}
\vspace{1.2cm}
\label{fig2--continued}
\end{figure}

\clearpage

\begin{figure}
\figurenum{2--continued}
\epsscale{0.40}
\vspace{0.7cm}
\plotone{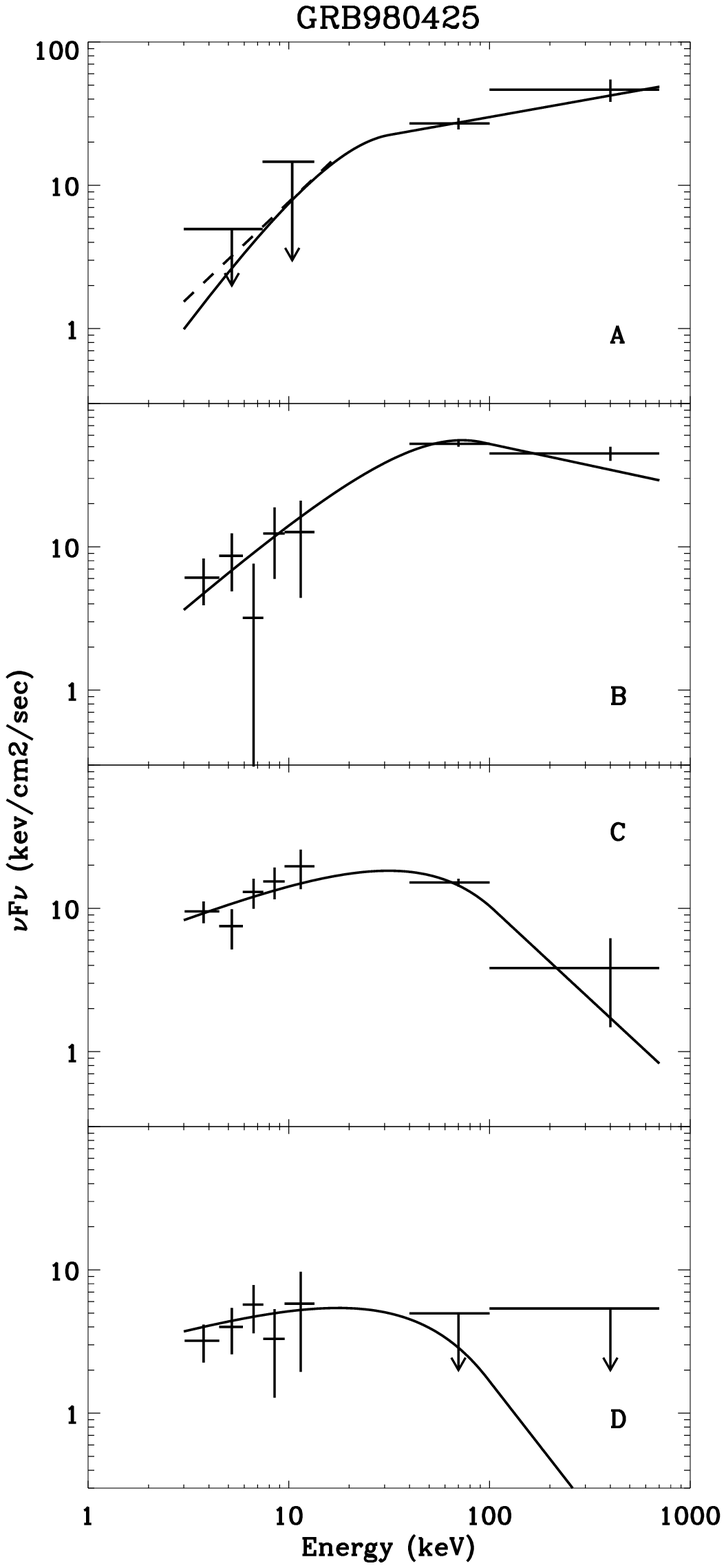}
\caption[]{}
\vspace{1.2cm}
\label{fig2--continued}
\end{figure}

\clearpage

\begin{figure}
\figurenum{3}
\epsscale{0.9}
\plotone{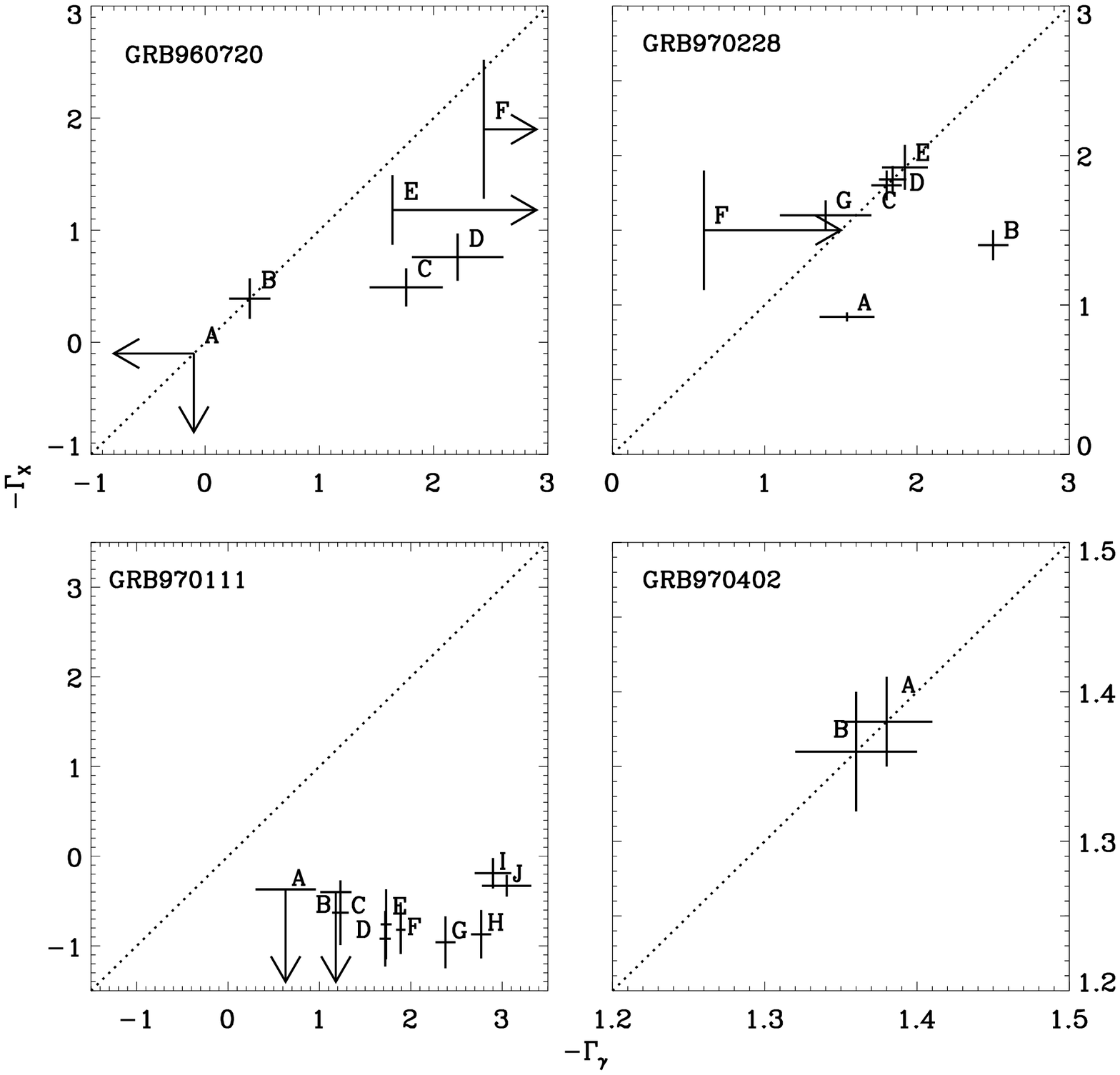}
\vspace{1.2cm}
\caption[fig3a.ps,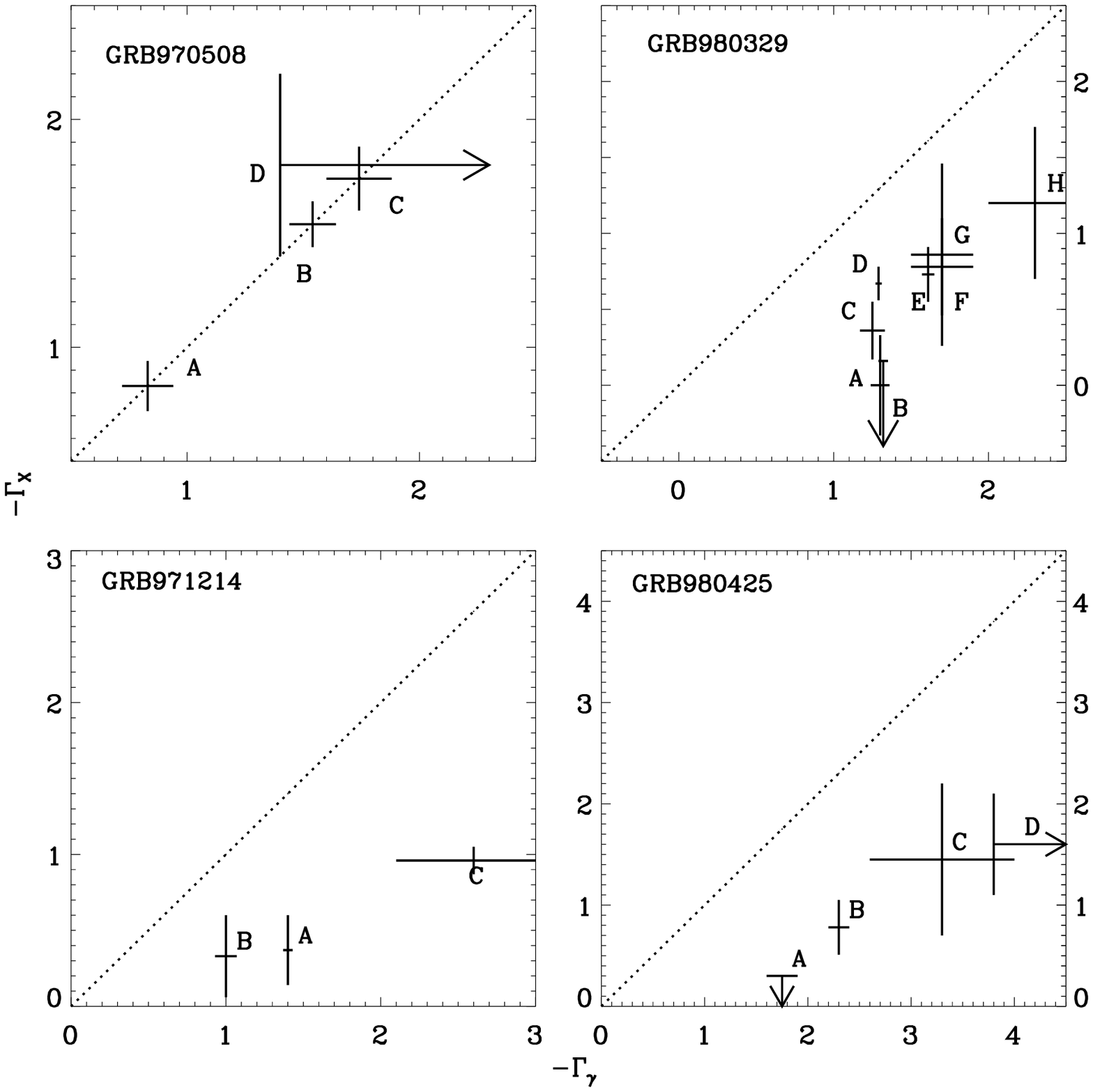]{Scatter plot of $\gamma$--ray photon index 
($\Gamma_{\gamma}$) vs. X--ray photon index ($\Gamma_{X}$). The dashed 
line shows the locus of $\Gamma_{\gamma}=\Gamma_{X}$ (see text).} 
\label{fig3}
\end{figure}

\clearpage

\begin{figure}
\figurenum{3--continued}
\epsscale{0.9}
\plotone{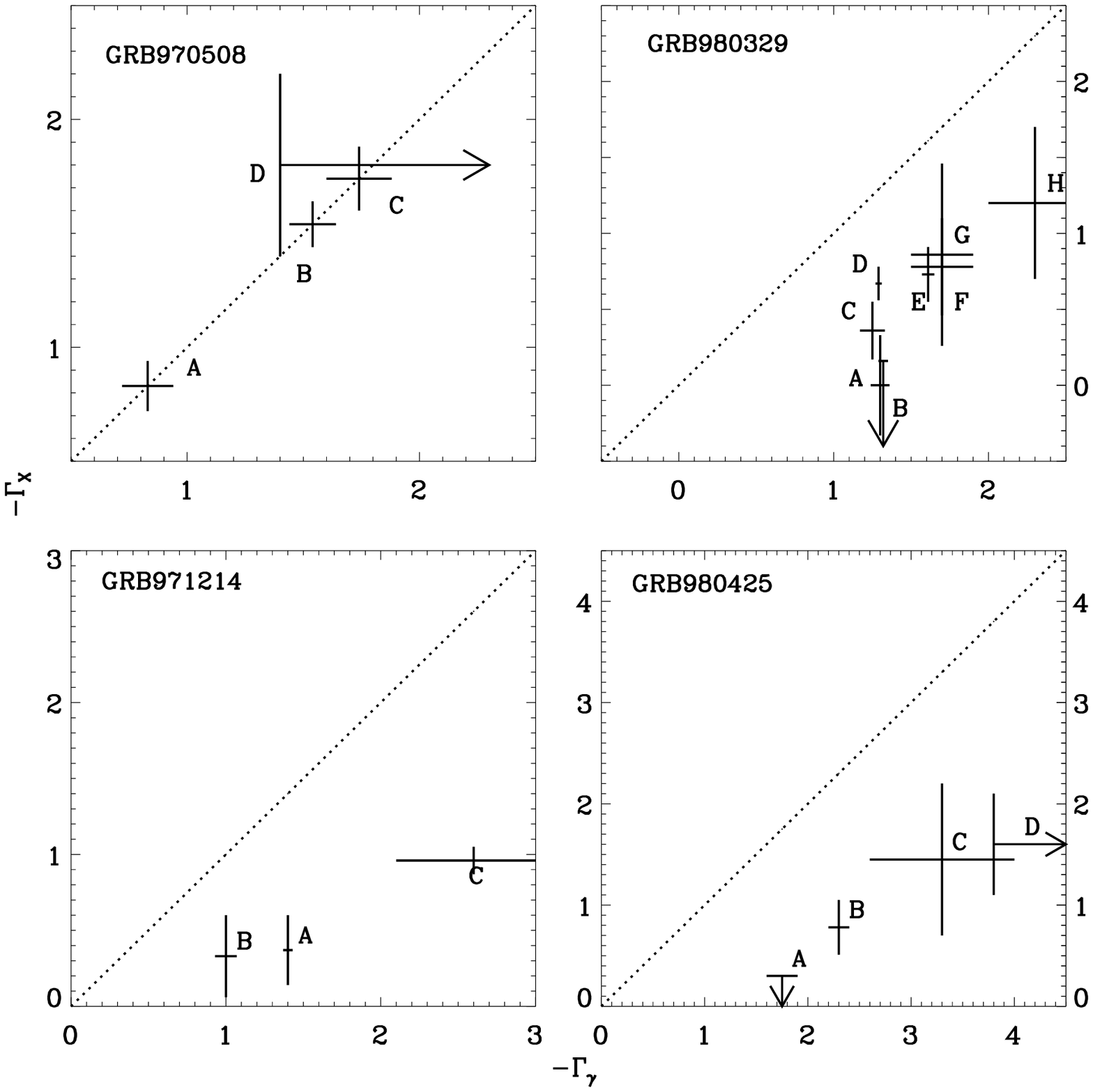}
\caption[]{}
\vspace{1.2cm}
\label{fig3--continued}
\end{figure}

\clearpage

\begin{figure}
\figurenum{4}
\epsscale{0.9}
\plotone{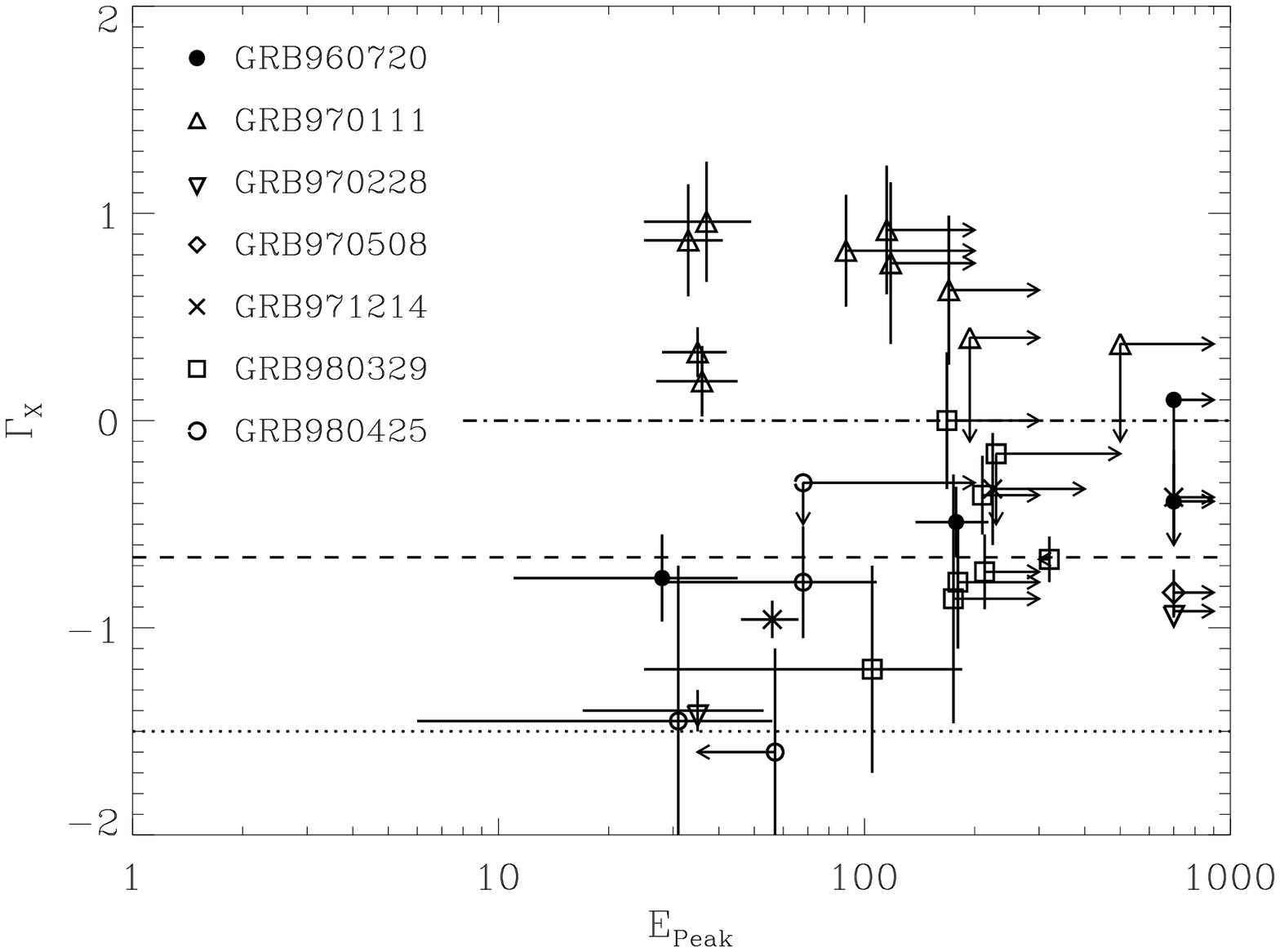}
\vspace{1.2cm}
\caption[fig4.ps]{Low-energy power law index $\Gamma_X$ vs. peak energy
E$_p$ of the $\nu$F($\nu$) spectrum, for each of the temporal sections of
GRBs in our sample. The dashed line corresponds to the maximum
power law photon index below the energy break, that is consistent with an 
optically thin synchrotron shock model, while the dotted line corresponds 
to the power law index when cooling of the high energy electrons is taken 
into account (see text). The dashed-dotted line correponds to the limit photon
index of the Inverse Compton (see text). 
}
\label{fig4}
\end{figure}

\clearpage

\begin{figure}
\figurenum{5}
\epsscale{0.9}
\plotone{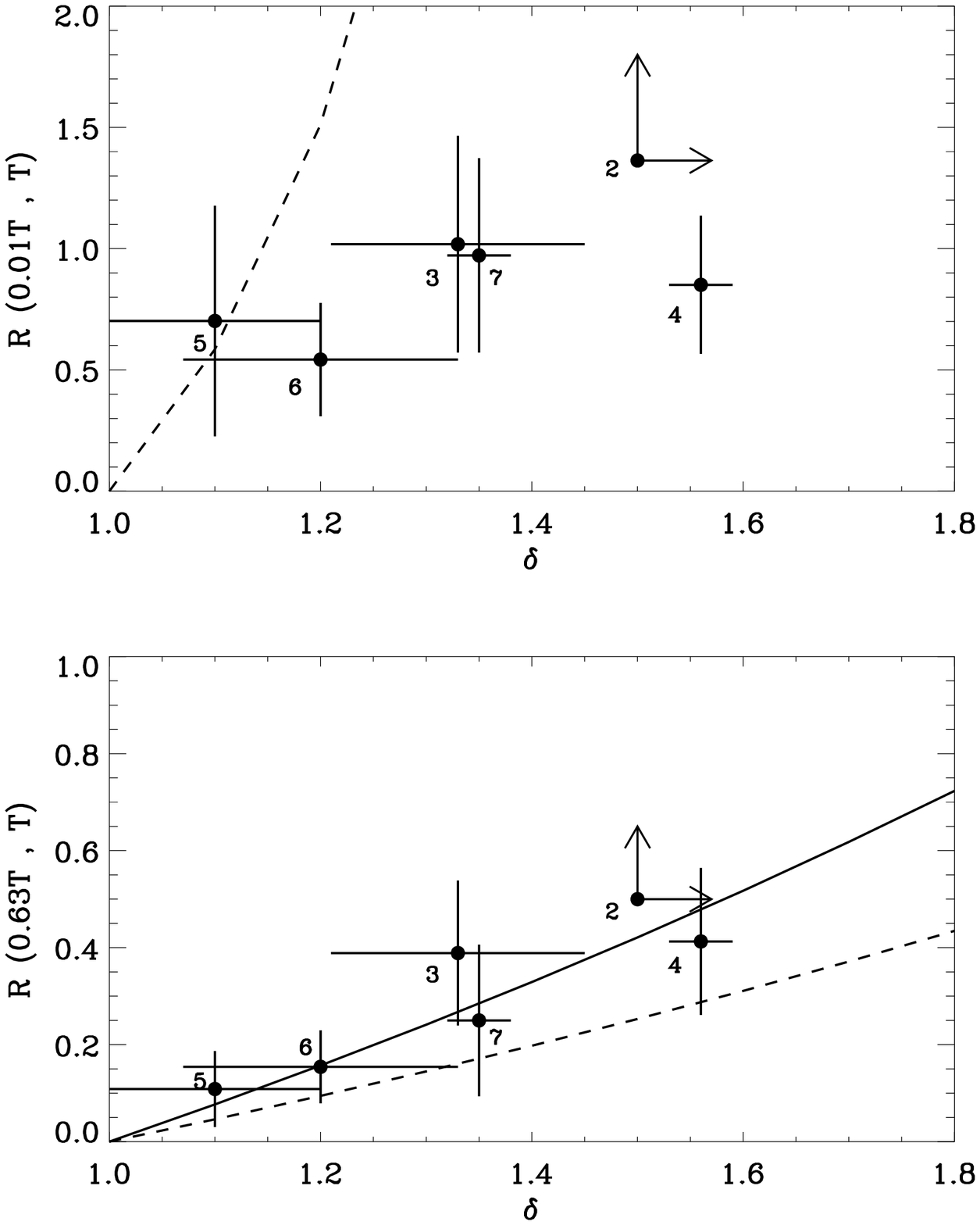}
\vspace{1.2cm}
\caption[fig5.ps]{Ratio between the 2-10~keV fluence derived
from the GRB light curve ({\it GRB fluence}) and that derived from the late 
afterglow observation ({\it afterglow fluence}) vs. the temporal index 
$\delta$ of the late afterglow fading law. 
The afterglow fluence is integrated from the end (T) of the GRB light curve, 
as estimated by our data (see Table~1) to 10$^6$~s (see text).\\ 
{\it Top}: The GRB fluence is integrated over the time interval starting
from 1\% of the GRB time duration to its end.\\
{\it Bottom}: The GRB fluence is integrated over the time interval 
starting from 63\% of the GRB time duration and the end of the event.
\\
Superposed to the data are the expected ratios  assuming the GRB fluence 
has the same origin as the late afterglow emission. 
{\it Continuous line}: best fit curve; {\it dashed line}:
expected curve if there is no spectral softening from the prompt to the
late afterglow emission. The best fit curve in the {\it top} case would
require a hardening of the spectra that is in contrast with the observations.
}

\label{fig5}
\end{figure}

\clearpage

\begin{figure}
\figurenum{6}
\epsscale{0.80}
\plotone{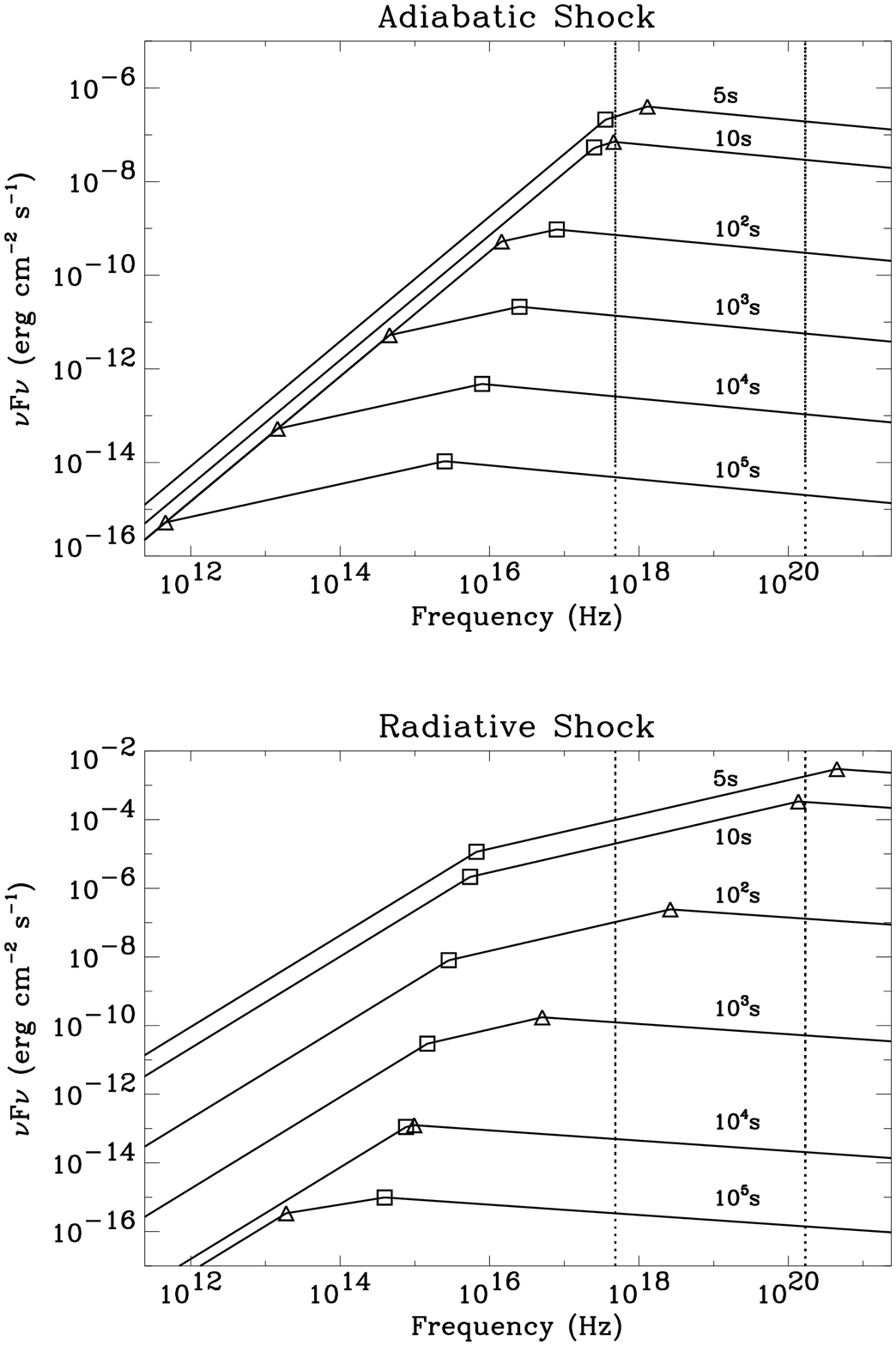}
\vspace{-0.4cm}
\caption[fig6.ps]{Theoretical afterglow spectra at various 
times from the afterglow onset, expected on the basis of the
synchrotron shock model by Sari, Piran and Narayan (1998). 
{\it Top}: a fully adiabatic shock; {\it bottom}: a fully radiative shock. 
Triangles correspond to the break energy E$_m$, while the squares correspond
to the cooling break E$_c$ (see text for definitions). The dotted vertical 
lines limit our energy passband. The parameters
assumed are: energy of the shock E~=~10$^{52}\,$erg, index of the power-law
distribution of the electrons p~=~2.3,
particle density of the interstellar medium n~=~1~cm$^{-3}$, fireball 
distance D~=~10$^{28} \,$cm, initial Lorentz factor of the ejecta 
$\gamma_0$~=100. The other parameters (see text) $\epsilon_e$ and 
$\epsilon_B$ depend on the type of cooling of the shock. We have assumed 
$\epsilon_B$~=~0.01 and $\epsilon_e$~=~0.1 for an {\it adiabatic shock},
$\epsilon_B$~=~0.1 and $\epsilon$$_e$~=~0.8 for a {\it radiative shock}.}

\label{fig6a}
\end{figure}

\clearpage


\begin{figure}
\figurenum{7}
\epsscale{0.9}
\plotone{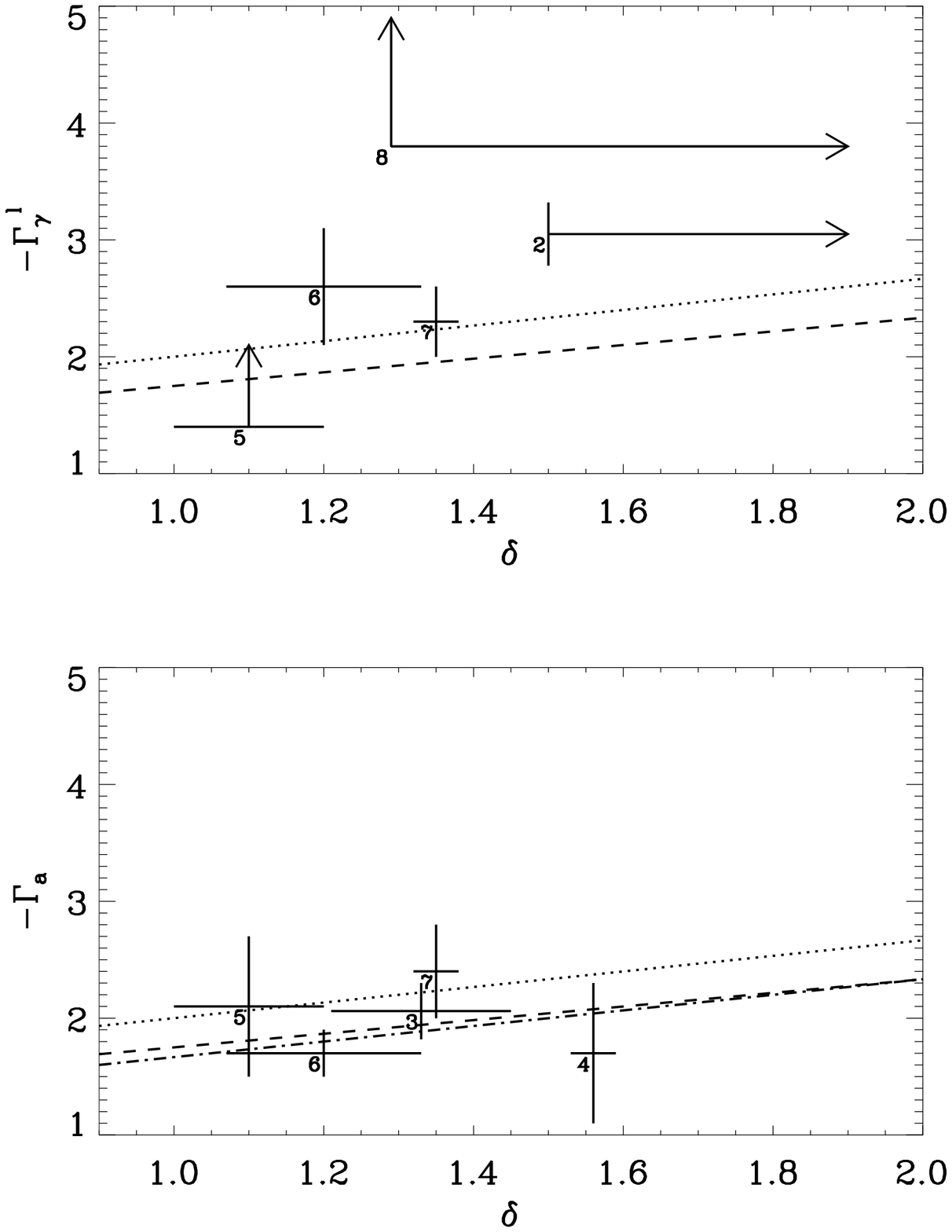}
\vspace{1.2cm}
\caption[fig7.ps]{ {\it Top panel:} High-energy power-law photon 
index $\Gamma_\gamma^l$ during the GRB tail vs. index of  the afterglow fading 
power-law. {\it Bottom panel:} photon index $\Gamma_a$ of the power law 
spectrum of the X-ray afterglow vs. index of  the afterglow fading 
power-law.  
{\it Dotted line}: expected relationship for a fully adiabatic 
shock if the  afterglow observation is performed at time t$>$t$_c$; 
{\it Dotted-dashed line}: expected relationship for a fully adiabatic 
shock if the afterglow observation is performed at time t, with  
t$_m<$t$<$t$_c$;  
{\it Dashed line}: expected relationship for a fully radiative shock. 
See also text.
}
\label{fig7}
\end{figure}

\clearpage

\begin{figure}
\figurenum{8}
\epsscale{0.82}
\plotone{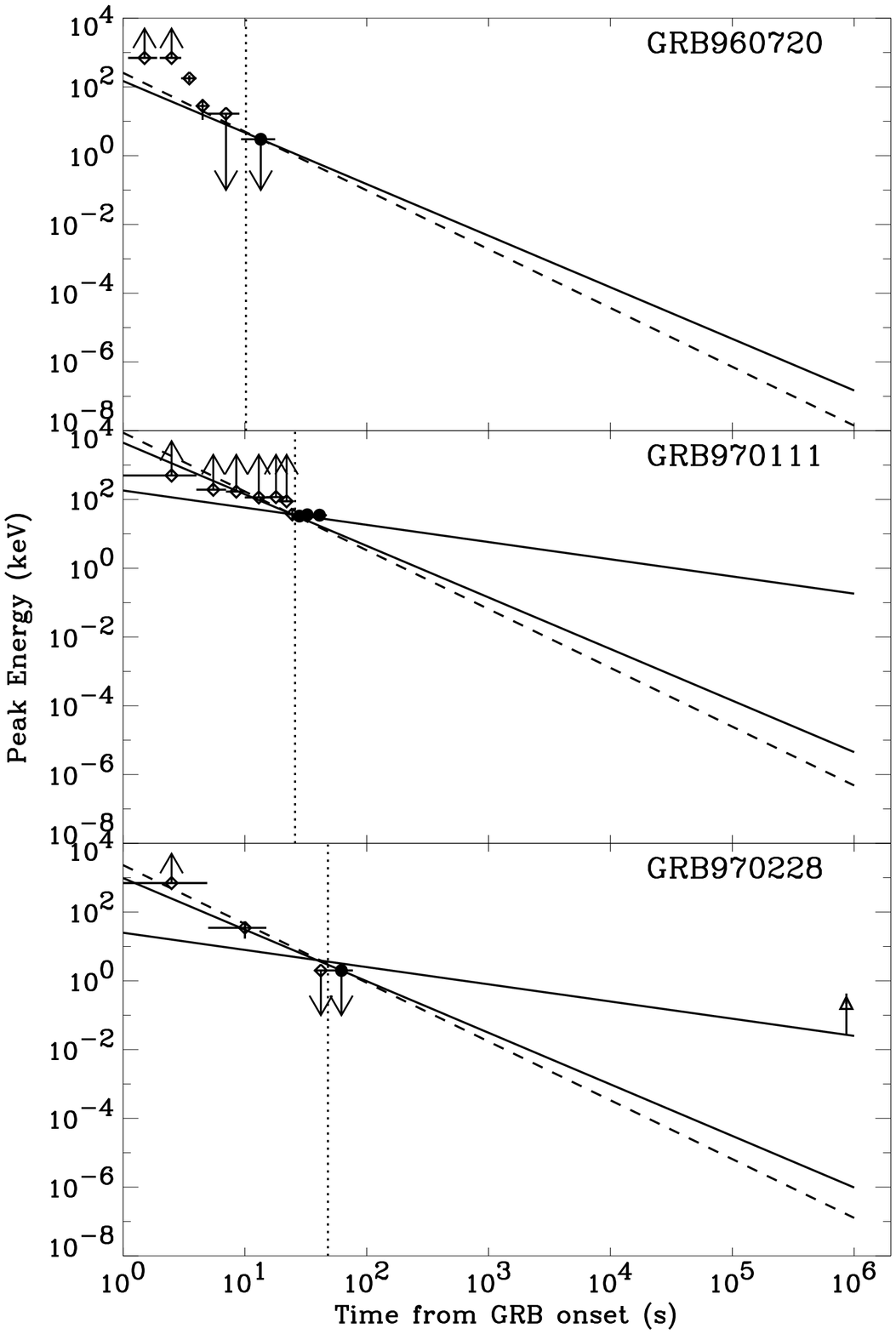}
\vspace{1.4cm}
\caption[fig8a.ps,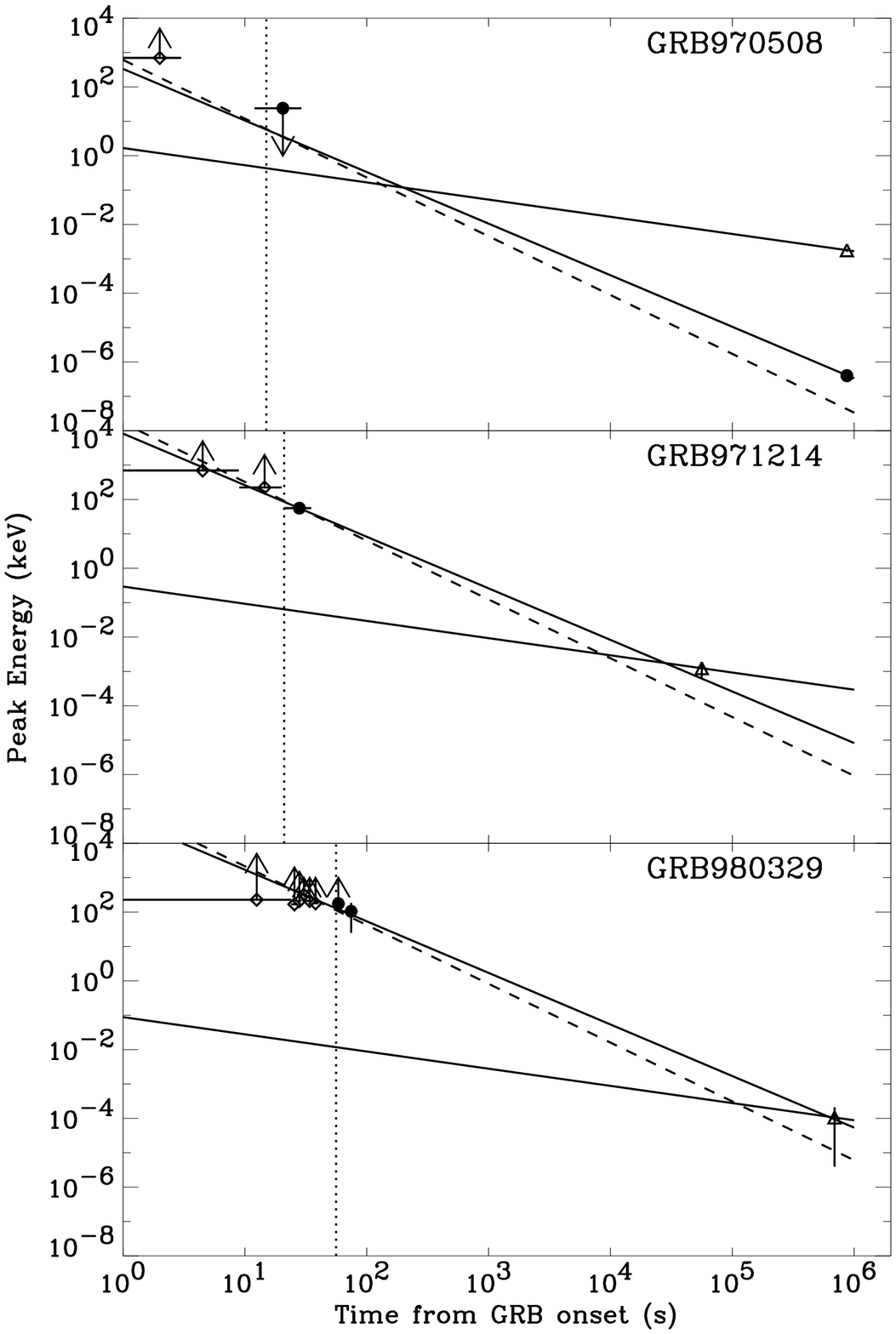,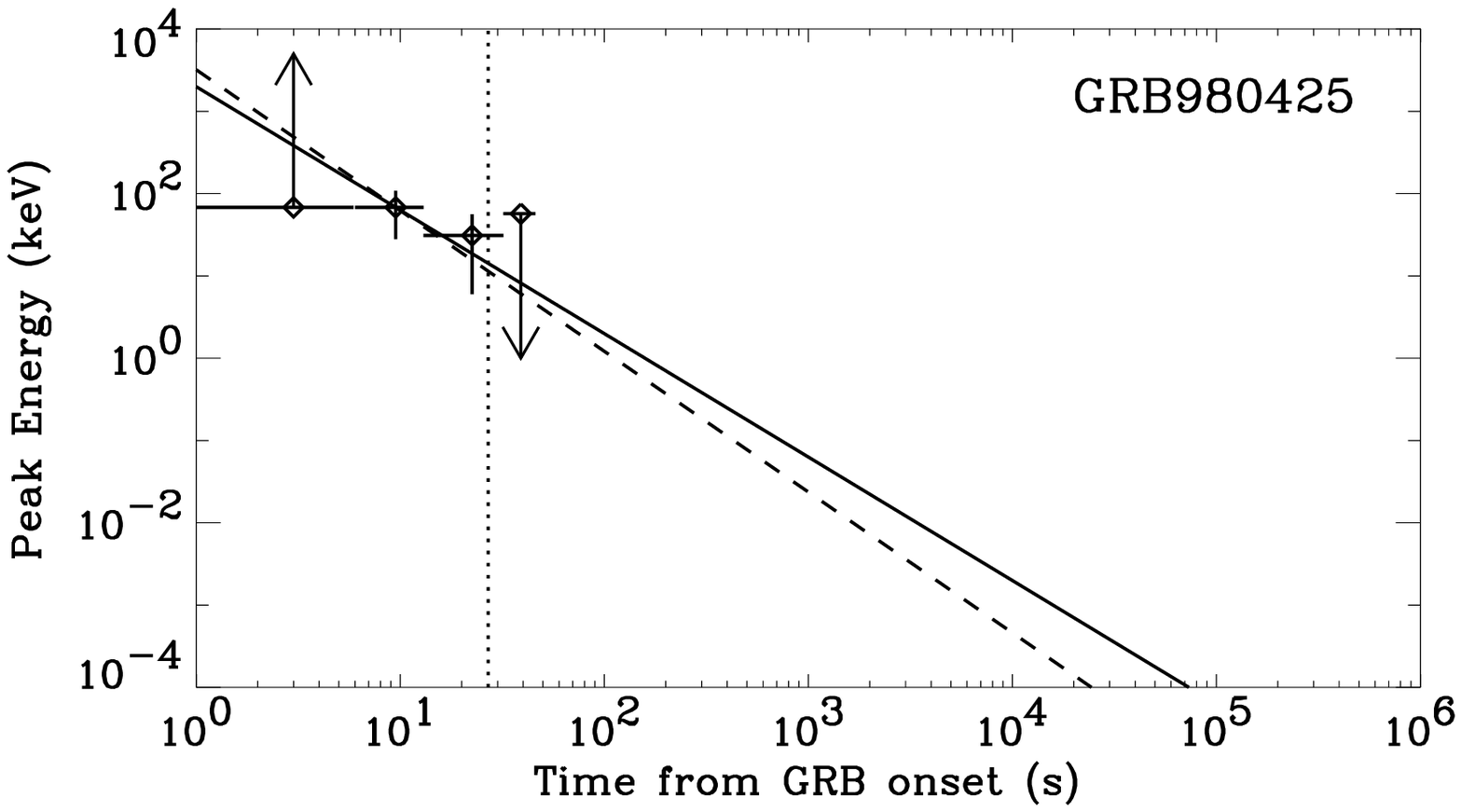]{Time behaviour of the measured peak 
energy E$_{p}$ of the $\nu$F($\nu$) spectrum for GRBs 
in our sample. {\it Vertical dotted line}: time at which early afterglow 
is expected to have already started (see Sect.\ref{afterglow}). 
E$_p$ values that are expected to be coincident with E$_m$ are shown as 
{\it full circles}, while those coincident with  E$_c$ are 
shown with {\it open triangles}. 
The expected time behaviour of E$_{p}$ in the case of the synchrotron shock 
model by Sari, Piran and Narayan (1998) is also shown for 
two solutions of the shock: adiabatic (continuum lines) and radiative
(dashed lines) cooling. Line with higher slope corresponds to fast cooling 
and that with lower slope corresponds to slow cooling. See text for details.
}
\label{fig8}
\end{figure}

\clearpage

\begin{figure}
\figurenum{8--continued}
\epsscale{0.82}
\plotone{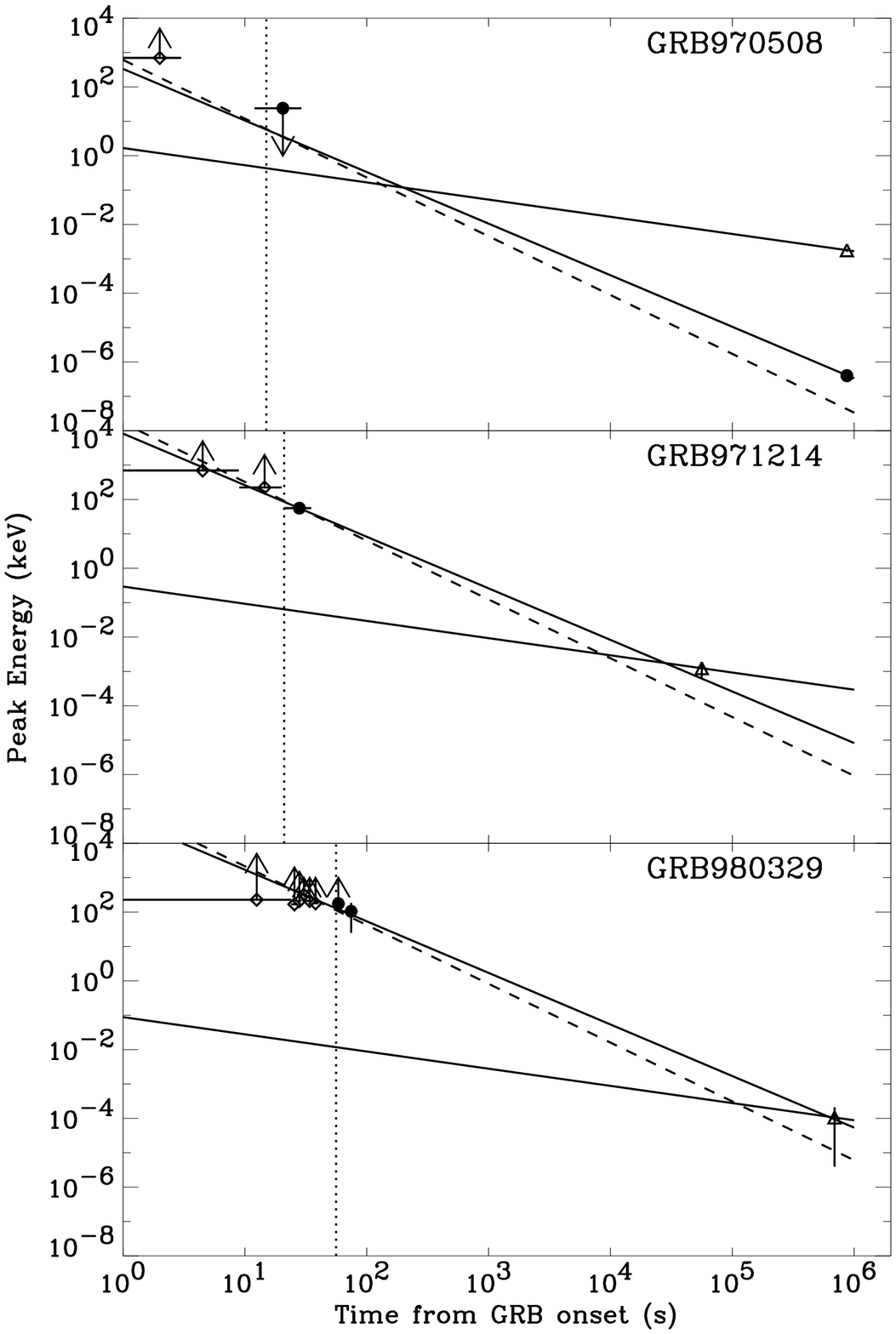}
\vspace{0.5cm}
\caption[]{}
\vspace{1.2cm}
\label{fig8--continued}
\end{figure}

\begin{figure}
\figurenum{8--continued}
\epsscale{0.82}
\plotone{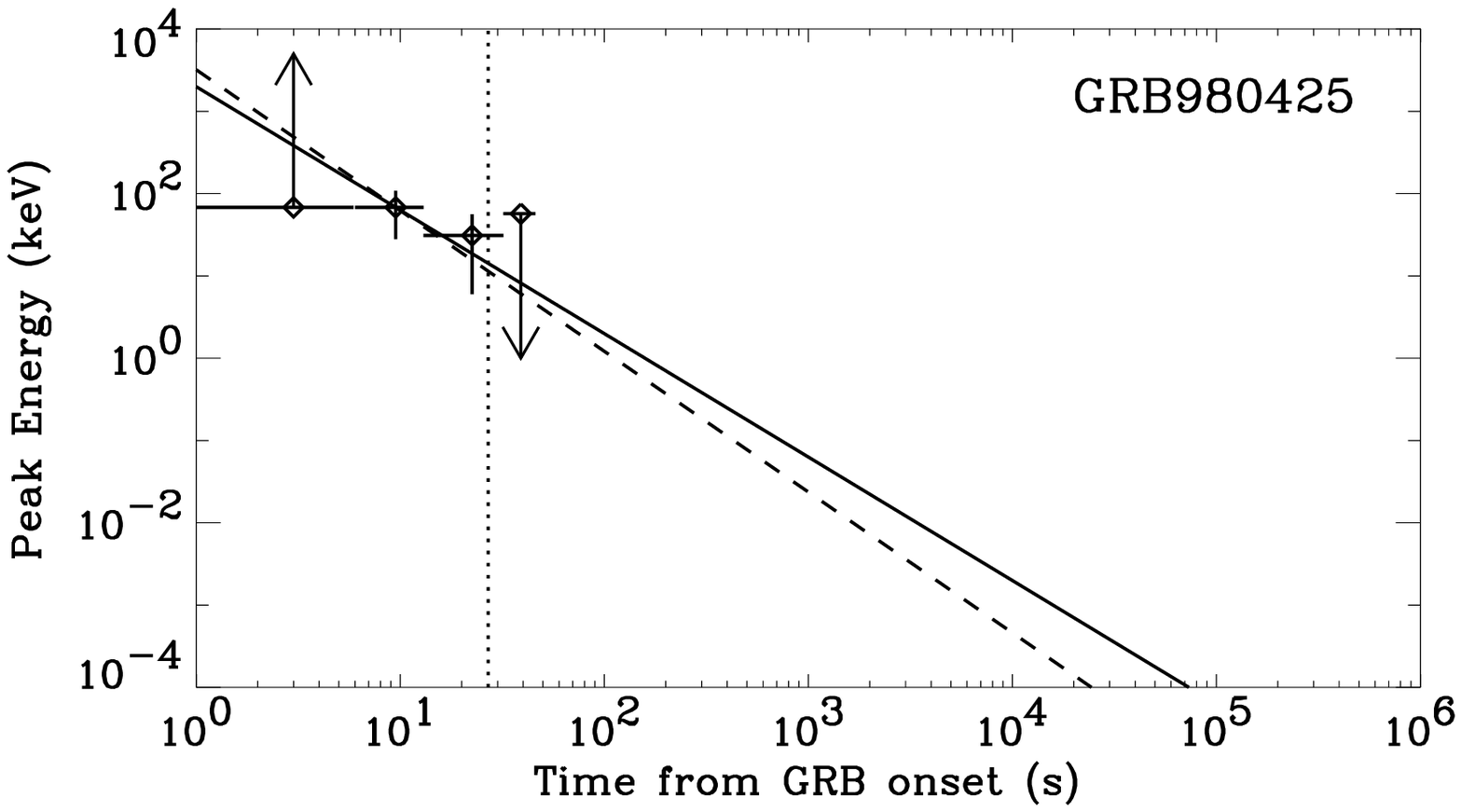}
\vspace{0.5cm}
\caption[]{}
\vspace{1.2cm}
\label{fig8--continued}
\end{figure}

\end{document}